\documentclass[apj,twocolappendix]{emulateapj}
\submitted{ApJ, in press}
\usepackage{hyperref}

\newcommand{\pwn}{G283.1$-$0.59}
\newcommand{\psr}{PSR~J1015$-$5719}
\newcommand{\kms}{\,km\,s$^{-1}$}
\newcommand{\ergs}{\,erg\,s$^{-1}$}

\newcommand{\mjb}{\,mJy\,beam$^{-1}$}
\newcommand{\ujb}{\,$\mu$Jy\,beam$^{-1}$}
\newcommand{\edot}{\mbox{$\dot{E}$}}

\newcommand\rs[1]{_\mathrm{#1}}

\newcommand{\tht}{\theta}
\newcommand{\tr}{\tilde r}
\newcommand{\thi}{\theta_{\rm int}} 
\newcommand{\ttht}{\tilde\theta}
\newcommand{\tthti}{\tilde\theta_{\rm int}}

\newcommand\rbs{r\rs{bs}}
\newcommand\aproj{a\rs{proj}}
\newcommand\rlos{r\rs{los}}
\newcommand\rproj{r\rs{proj}}
\newcommand\xlos{x\rs{los}}

\hypersetup{colorlinks, citecolor=blue, filecolor=blue, pdfborderstyle={},
linkcolor=blue, urlcolor=blue, pdfpagemode=UseNone,
pdfborder={0 0 0}}

\begin{document}
\title{Discovery of a Synchrotron Bubble Associated with \psr}

\shorttitle{A Synchrotron Bubble Powered by \psr}
\shortauthors{Ng et al.}

\author{C.-Y.~Ng\altaffilmark{1}, R.~Bandiera\altaffilmark{2},
R.~W.~Hunstead\altaffilmark{3}, and S.~Johnston\altaffilmark{4}}
\affiliation{$^1$Department of Physics, The University of Hong Kong, Pokfulam
Road, Hong Kong; \url{ncy@bohr.physics.hku.hk} }
\affiliation{$^2$Osservatorio Astrofisico di Arcetri, Largo E. Fermi 5, I-50125 Firenze, Italy}
\affiliation{$^3$Sydney Institute for Astronomy (SIfA), School of Physics, The University of Sydney, NSW 2006, Australia}
\affiliation{$^4$CSIRO Astronomy and Space Science, P.O. Box 76, Epping NSW 1710, Australia}



\begin{abstract}
We report the discovery of a synchrotron nebula, \pwn, associated with \psr.
Radio observations using the Molonglo Observatory Synthesis Telescope and the
Australia Telescope Compact Array at 36, 16, 6, and 3 cm reveal a complex
morphology. The pulsar is embedded in the ``head'' of the nebula with
fan-shaped diffuse emission. This is connected to a circular bubble of
20\arcsec\ radius and a collimated tail extending over 1\arcmin. Polarization
measurements show a highly ordered magnetic field in the nebula. It wraps
around the edge of the head and shows an azimuthal configuration near the
pulsar, then switches direction quasi-periodically near the bubble and in the
tail. Together with the flat radio spectrum observed, we suggest that this
system is most plausibly a pulsar wind nebula (PWN), with the head as a bow
shock that has a low Mach number and the bubble as a shell expanding in a
dense environment. The bubble could act as a magnetic bottle trapping the
relativistic particles. A comparison with other bow-shock PWNe with higher
Mach numbers shows similar structure and $B$-field geometry, implying that
pulsar velocity may not be the most critical factor in determining the
properties of these systems. 

We also derive analytic expressions for the projected standoff distance and
shape of an inclined bow shock. It is found that the projected distance is
always larger than the true distance in three dimensions. On the other hand,
the projected shape is not sensitive to the inclination after rescaling with
the projected standoff distance. 
\end{abstract}

\section{\bf Introduction}
As a pulsar spins down, most of its rotational energy is carried away by a
relativistic particle and magnetic field outflows known as a pulsar wind. The
interaction of the wind with the ambient medium results in a pulsar wind
nebula (PWN), emitting broadband synchrotron radiation from radio to X-ray
bands. As an aged pulsar generally travels in the interstellar medium
(ISM) faster than the local sound speed, the wind is confined by the ram
pressure, giving rise to a bow-shock PWN. There are over a dozen of such
systems known and most of them are moving at a high Mach number
\citep[see][]{kp08}. They are characterized by a long collimated tail with a
small bow-shock standoff distance, which makes them easier to identify than
low-velocity examples.

While bow shocks are governed by simple boundary conditions, they exhibit
diverse properties. Observations reveal a peculiar structure with a bubble or
a central bulge \citep[e.g.][]{crl93,ngc+10} and different magnetic field
geometry \citep[e.g.,][]{yg05,ngc+10,nbg+12}. The exact origin of these
features remains unclear. It could be related to the pulsar age, velocity,
flow condition of the wind, the relative orientation of the pulsar spin, etc.
To identify which factor plays a critical role, it is essential to
expand the sample. We present here the discovery of a new bow-shock PWN,
\object{\pwn}, associated with \object{\psr} (hereafter, J1015). As we argue,
this is an unusual system traveling at a low Mach number and showing many
remarkable properties, thus offering an important example for understanding
bow shocks.

\begin{deluxetable*}{llrrrcc}
\tablewidth{0pt}
\tablecaption{Radio Observations of J1015 Used in This Study \label{table:obs}}
\tablehead{\colhead{Obs.\ Date} & \colhead{Array}&
\colhead{Wavelength}& \colhead{Center Freq.} &
\colhead{No.\ of} & \colhead{Usable Band-} & \colhead{Integration} \\
& \colhead{Config.} & \colhead{(cm)} & \colhead{(MHz)} &
\colhead{Channels\tablenotemark{a}} & 
\colhead{\protect{width\tablenotemark{a}} (MHz)} 
& \colhead{Time (hr)}}
\startdata
\multicolumn{5}{l}{MOST}\\
2008 Apr 6, 13, 19, 20 & \nodata & 36& 843 & 1 & 3 & 48 \\
\multicolumn{5}{l}{ATCA}\\
2008 Dec 13 & 750B & 6, 3 & 4800, 8640 & 13 & 104 & 12.5 \\
2009 Feb 17 & EW352 & 6, 3 & 4800, 8640 & 13 & 104 & 12.5 \\
2009 Aug 2 & 1.5A & 6, 3 & 5500, 9000 & 2048 & 1848 & 11.5 \\
2010 Feb 5 & 6A & 6, 3 & 5500, 9000 &  2048 & 1848 & 10 \\
2011 Nov 17 & 1.5D & 16 & 2100, 2102\tablenotemark{b}& 2048,
256\tablenotemark{b}& 1848 & 1 \\
\enddata
\tablenotetext{a}{Per center frequency.}
\tablenotetext{b}{Taken with the pulsar binning mode.}
\end{deluxetable*}
The MOST data reduction followed the custom procedure outlined in \citet{bls99}
and \citet{gcl+99}. We first removed those data samples most strongly affected
by radio frequency interference (RFI), then calibrated the pointing and flux
density scale using sources listed in \citet{ch94}. Finally, we formed an
intensity map and performed deconvolution with a CLEAN algorithm \citep{hog74}.
This was straightforward because of the nearly continuous $u$--$v$ coverage of
the MOST observation. The final radio image at 843\,MHz was formed using the
task \texttt{IMCOMB} in MIRIAD \citep{stw95} to co-add the four
separate images; the beam size was $43\arcsec\times 51\arcsec$ FWHM and the rms
noise in regions away from the diffuse emission was $\sim2$\mjb.

J1015 is an energetic pulsar discovered in the Parkes Multibeam Pulsar Survey
\citep{kbm+03}. It has spin period $P=0.14$\,s, spin-down luminosity $\dot
E=8.3\times10^{35}$\ergs, and characteristic age $\tau_c=39$\,kyr. The
dispersion measure (DM) gives a distance estimate of 5.1\,kpc \citep{cl02}.
\citet{jw06} performed polarization measurements of the pulsar, and found that
the swing of the position angle (PA) across phase can be well fitted by the
rotating vector model. They inferred a PA of 62\arcdeg\ (north through east)
for the projected pulsar spin axis on the plane of the sky. The pulsar is
located near the gamma-ray source 3EG~J1014$-$5705 detected with EGRET.
Although an association had been suggested \citep{tbc01}, recent \emph{Fermi}
LAT observations identified the gamma-ray source as 3FGL~J1013.6$-$5734 and
the refined position is over 20\arcmin\ from J1015 \citep{aaa+15}. 
J1015 now lies beyond the 95\% error ellipse of the gamma-ray
position, making the association unlikely. So far there is no detection of
pulsed or persistent gamma-ray emission at the pulsar position. Infrared,
optical, and X-ray observations also found no counterparts \citep{wnw+14}.

A radio image of the field from the second epoch Molonglo Galactic Plane
Survey \citep[MGPS2;][]{mmg+07,grm14} shows extended emission near J1015 at
36\,cm. In this study we present deeper radio observations with the Molonglo
Observatory Synthesis Telescope (MOST), and higher-resolution observations
taken with the Australia Telescope Compact Array (ATCA). The observations and
data reduction are described in Section~\ref{sec:obs}. The results are
reported in Section~\ref{sec:result} and are discussed in
Section~\ref{sec:discuss}. We summarize the findings in
Section~\ref{sec:concl}. Finally, we derive analytic expressions of the
projected standoff distance and shape of an inclined bow shock in
Appendix~\ref{sec:app_bs}, and we present an evolution model of a bubble with
continuous energy injection in a uniform medium in Appendix~\ref{sec:app_snr}.
The new images resolved the extended emission and revealed a bubble-like
structure, with a high degree of linear polarization and a flat spectrum. We
suggest that this is a newly identified bow-shock PWN associated with J1015.

\section{\bf Observations and Data Reduction}
\label{sec:obs}
Radio continuum observations of the field of J1015 were made with MOST and
ATCA, and the details are listed in Table~\ref{table:obs}. MOST
operated at a wavelength of 36\,cm with a bandwidth of 3\,MHz, and it only
recorded the right-hand circular polarization. For ATCA, J1015 was observed at
3 and 6\,cm with a 4-pointing mosaic in different array configurations. Two
early observations had a useful bandwidth of $\sim$100\,MHz, and two later
ones were taken after the compact array broadband backend upgrade
\citep{wfa+11}, which provided a much larger bandwidth of 2\,GHz. Further data
obtained on 2009 August 8 were not usable due to an instrumental problem. At
16\,cm, the ATCA observation was part of a snapshot survey, and the total
integration time on J1015 was 1\,hr spread over 12\,hr. It was done
simultaneously in both the standard observing mode and the pulsar binning
mode. The latter provided a high time resolution and the pulse period of J1015
was divided into 32 phase bins to measure the pulse profile and to form on-
and off-pulsed images. All ATCA data had full Stokes parameters recorded.
PKS~B1934$-$638 was used as the flux and bandpass calibrator. PKS~B1036$-$52
was used as the phase calibrator at 3 and 6\,cm, while PKS~B1049$-$53 was used
at 16\,cm.

We carried out the ATCA data reduction and analysis using MIRIAD. We first
removed edge channels and flagged bad visibility data points affected by RFI,
then employed the standard procedure to determine and apply the bandpass,
gain, flux, and polarization calibration corrections. The 3 and 6\,cm images
were formed with natural weighting, multifrequency synthesis, and a Gaussian
taper of FWHM 4\arcsec\ to boost the signal-to-noise ratio (S/N). We
deconvolved the mosaicked images in Stokes \emph{I}, \emph{Q}, and \emph{U}
jointly using a maximum entropy algorithm \citep[\texttt{PMOSMEM;}][]{sbd99}
and restored them with a circular Gaussian beam of FWHM 4\arcsec. The final 3
and 6\,cm maps had rms noise of 20\ujb\ and 15\ujb, respectively, consistent
with the theoretical values.

For the 16\,cm observation, we applied the same flagging and calibration
procedures as above. The pulsar binning data were corrected for dispersion
using the DM value for J1015. Since the fractional bandwidth is very large, we
divided the data into eight bands of 256\,MHz to form separate images. The
\citet{bri95} weighting scheme was employed with the \texttt{robust} parameter
of 0.5, which optimizes the weighting between resolution and noise level.
Images were deconvolved using the CLEAN algorithm and restored with a gaussian
beam. The results were combined to form final images in Stokes \emph{I},
\emph{Q}, and \emph{U}. After averaging, the center frequency was 2.2\,GHz and
the images had beams of FWHM $8\arcsec\times7 \arcsec$ and rms noise of
50\ujb.

\begin{figure*}[ht]
\centering
\epsscale{1.0}
\plottwo{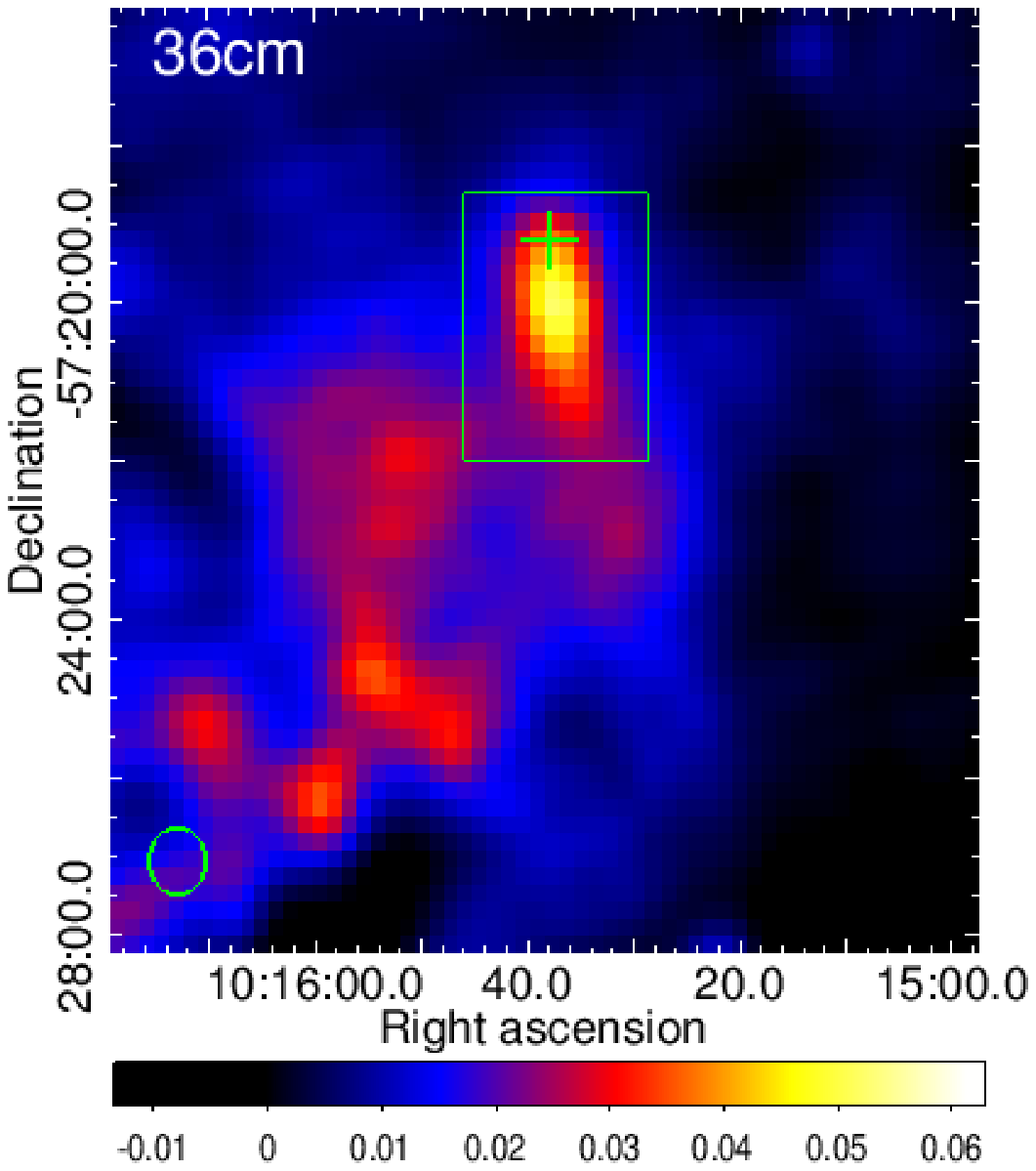}{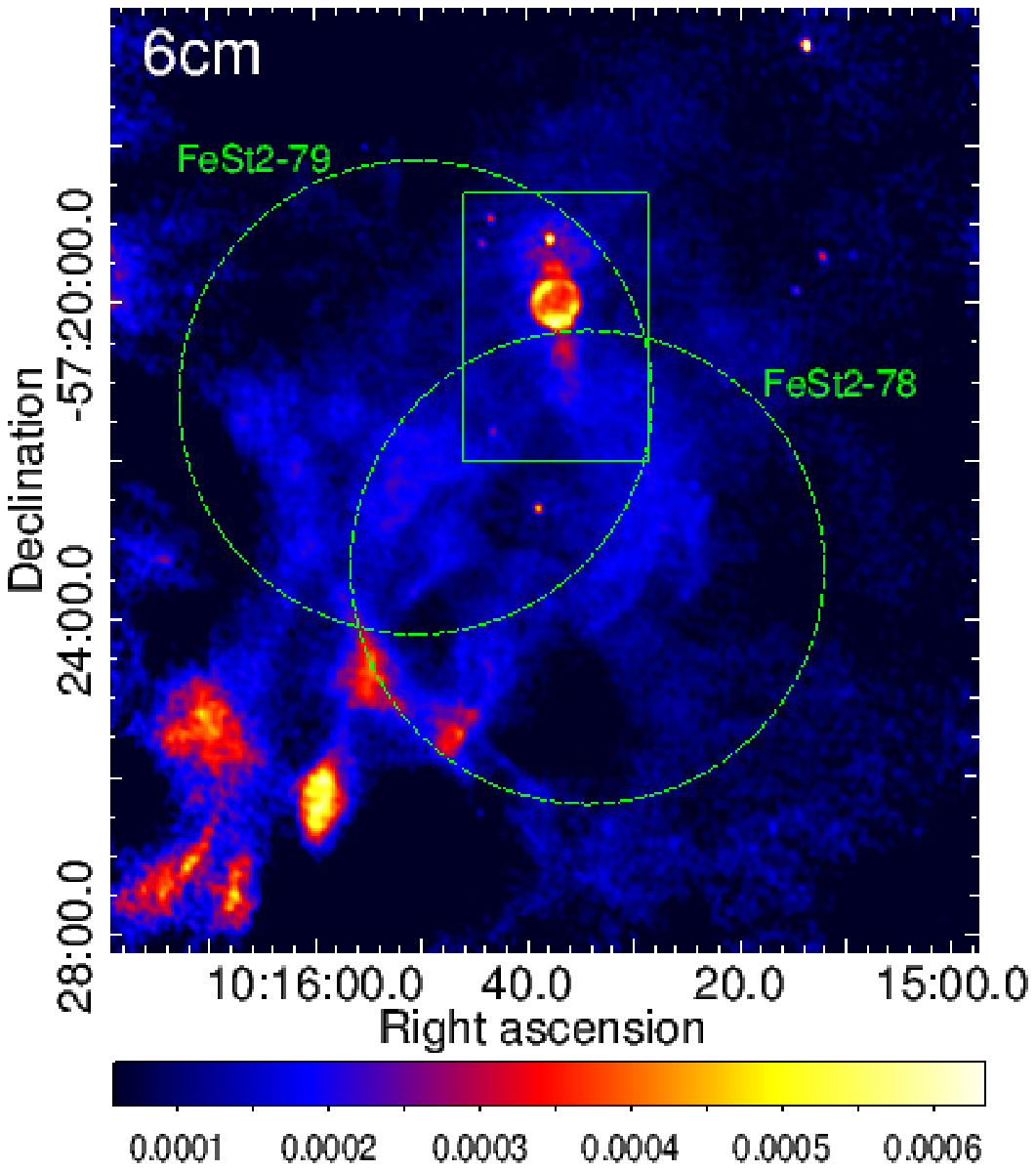}
\caption{Radio intensity maps of the field of J1015 at 36 and 6\,cm taken
with MOST and ATCA, respectively. In the MOST image, the cross marks the
pulsar position and the beam size is shown in the lower left. In the ATCA
image, the dashed circles indicate the location of two optical dust clouds,
FeSt2-78 and FeSt2-79 \citep{fs84,db02}. The circle diameters of 6\arcmin\
illustrate the resolution of the optical survey and the actual extents of the
clouds are likely smaller. The boxes show the field of view of
Figure~\ref{fig:zoom}. The color bars are in units of Jy\,beam$^{-1}$.
\label{fig:field}}
\end{figure*} 

\section{\bf Results}
\label{sec:result}
\subsection{Morphology}
\begin{figure*}[!th]
\centering
\epsscale{1.15}
\plotone{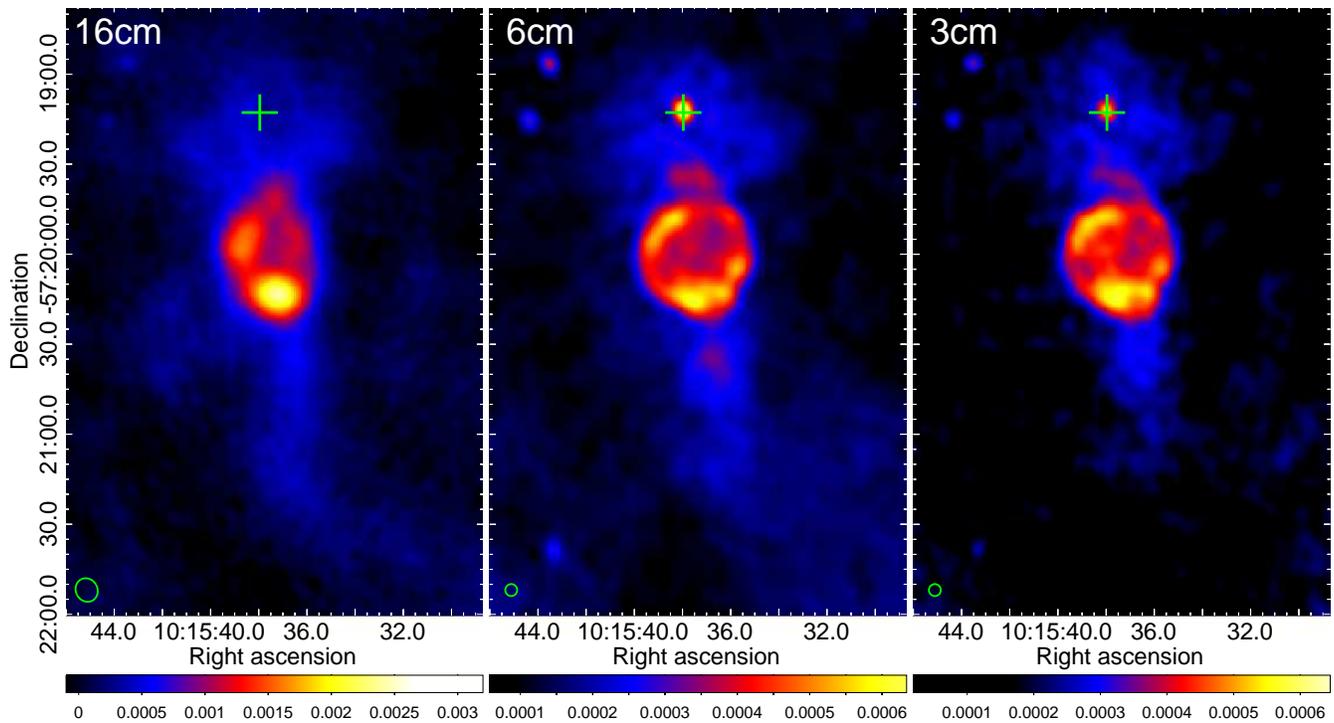}
\caption{ATCA radio intensity maps zoomed in at J1015 and the nebula \pwn\ at
16, 6, and 3\,cm. The 16\,cm image is obtained from the off-pulse phase bins
with the pulsar binning data (see Figure~\ref{fig:profile} for the pulse
profile of J1015.) The crosses mark the pulsar position and the beam sizes
are shown in the lower left. The color bars are in units of Jy\,beam$^{-1}$.
 \label{fig:zoom}}
\end{figure*}

Figure~\ref{fig:field} shows the radio intensity maps of the field of J1015.
The MOST 36\,cm image indicates a compact nebula, which we designate \pwn,
extending $\sim$2\farcm5 south from the pulsar position. Further southeast,
there is large-scale diffuse emission over 4\arcmin\ in size. The compact
nebula is resolved by the higher-resolution ATCA images at 16, 6, and 3\,cm in
Figure~\ref{fig:zoom}, revealing a complex morphology. J1015 is detected in
all the bands. It is surrounded by faint fan-shaped emission that we refer to
as the ``head'' of the nebula. The head is $\sim$1\farcm5\ wide and its apex
is 20\arcsec\ north of the pulsar. It connects to a circular bubble-like
structure in the south. The latter has a radius of 20\arcsec\ and the center
is $\sim$50\arcsec\ away from the pulsar. Beyond the bubble, there is a faint
collimated tail extending 1\arcmin\ further south, such that the entire nebula
elongates in the north--south direction with a PA of $\sim$185\arcdeg.

\begin{figure}[!ht]
\epsscale{1.0}
\plotone{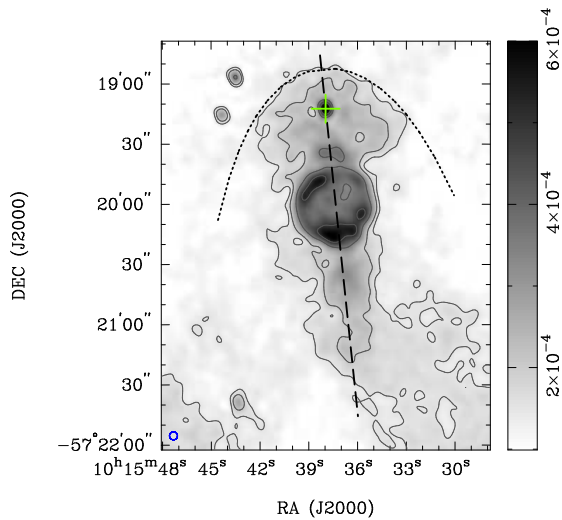}
\caption{The same 6\,cm intensity map of J1015 and \pwn\ as in
Figure~\ref{fig:zoom}, overlaid with intensity contours at levels of 0.15,
0.2, 0.35, and 0.5\mjb. The dotted curve shows a theoretical model of a
classical bow shock \citep{wil96} and the dashed line indicates its axis of
symmetry. The cross marks the pulsar position and the beam size is shown in
the lower left. The gray scale is linear and the color bar is in units of
Jy\,beam$^{-1}$. \label{fig:wilkin}}
\end{figure}

The morphology of the head resembles the shape of a classical bow shock. We
therefore compared the image with a theoretical model given by \citet{wil96}.
The result is plotted in Figure~\ref{fig:wilkin}, which indicates a very good
match and suggests a projected stand-off distance of $\sim$20\arcsec. The
bubble exhibits a similar morphology in all ATCA images\footnote{We note
that the bubble appears less circular at 16\,cm, but this is likely due to
poor image reconstruction resulting from the short integration time and hence
limited $u$-$v$ sampling.} and is much brighter around the rim than in the
interior. The emission peaks in the south and northeast rim with comparable
flux density at 6 and 3\,cm, while the northeast rim is fainter than the south
one at 16\,cm. The bright emission at the south rim seems to extend further to
the southwest in the 6 and 3\,cm images, although less obvious at 16\,cm.

\begin{figure}[th]
\epsscale{1.0}
\plotone{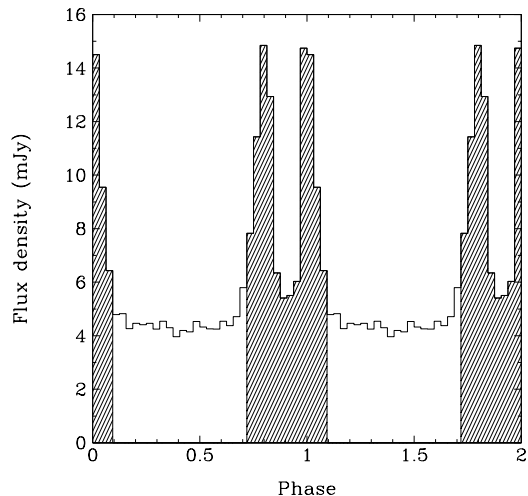}
\caption{Pulse profile of J1015 obtained with the 16\,cm pulsar binning
observation. Two cycles are plotted and the shaded and unshaded regions denote
the on- and off-pulse phases, respectively, chosen for the pulsed flux density
measurement. An image formed with the off-pulse phase is shown in
Figure~\ref{fig:zoom}.
\label{fig:profile}}
\end{figure}

We examined the infrared, optical, and X-ray images of the field using the
same data as in \citet{wnw+14}, but found no counterparts of \pwn.
There are two optical dark clouds, FeSt2-78 and FeSt2-79 \citep{fs84}
overlapping with the nebula. Their locations are shown in
Figure~\ref{fig:field} and their reported sizes are no larger than the survey
resolution of 6\arcmin\ \citep{db02}.
We attempted to identify the clouds in CO and H\,{\sc i} surveys
\citep{dht01,bfk+16}, but the resolution of these radio maps was too low
to be useful. Finally, in Figure~\ref{fig:profile} we plot the pulse profile
of J1015 obtained from the 16\,cm pulsar binning data. It exhibits a double
peak feature similar to that at 21\,cm \citep{jw06}, while the two peaks have
comparable intensity. Using the off-pulse data between phase 0.11 and 0.70, we
formed an intensity map of the nebula and it is shown in
Figure~\ref{fig:zoom}.

\subsection{Spectroscopy}
\begin{figure}[!ht]
\epsscale{1.0}
\plotone{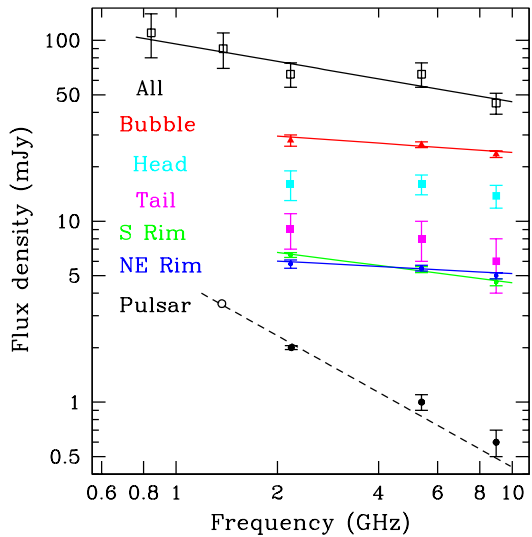}
\caption{Radio spectrum of J1015 and different regions of \pwn\ as listed in
Table~\ref{table:flux}. The open circle is the pulsed flux obtained with
single-dish observations \citep{jw06}. We do not show the fits for the head and
tail here due to the large uncertainties.\label{fig:spec}}
\end{figure}

We estimated the flux densities of \pwn\ from our radio maps and the Southern
Galactic Plane Survey \citep[SGPS;][]{hgm+06} 21\,cm image. The
lower-resolution MOST and SGPS images only allow us to measure the total flux
density of the entire nebula including the pulsar, while the higher-resolution
ATCA maps provide flux density measurements of different regions. At 16\,cm we
obtained the nebula flux density from the standard mode observation since it
has more channels than the pulsar binning data, thus better rejects the RFI
and gives a higher S/N image. The pulsar binning data were used to measure the
pulsar flux density, by subtracting the off-pulsed data from the on-pulsed
ones using the MIRIAD task \texttt{psrbl} (see Figure~\ref{fig:profile} for
the definition of the phase ranges). All results are listed in
Table~\ref{table:flux} and plotted in Figure~\ref{fig:spec}. The uncertainties
are mostly due to strong variations of the background. We also list the pulsar
flux density measured with single-dish observations for comparison. Note that
the updated value of 3.5\,mJy at 21\,cm \citep{jw06} is adopted here, instead
of $0.9\pm0.1$\,mJy reported in the discovery paper \citep{kbm+03}.

The spectral indices of different regions are determined from simple fits with a
power law $S_\nu\propto \nu^\alpha$. The overall spectrum of the nebula plus
pulsar is rather flat with an index $\alpha=-0.3$, while that of the pulsar is
much steeper ($\alpha=-1.04$) and that of the bubble is flatter
($\alpha=-0.13$). The south rim exhibits a steeper spectrum than the
northeast one ($\alpha=-0.24$ vs.\ $-0.10$). The head and tail generally
show flat spectra, albeit with large uncertainties since these features are
faint. We attempted to generate a spectral map of the field, but the S/N was
too low to be useful.

\begin{deluxetable*}{lcccr@{$\pm$}lr@{$\pm$}lr@{$\pm$}lcr@{$\pm$}l}[ht]
\tablewidth{0pt}
\tablecaption{Background-subtracted Flux Densities and Best-fit Spectral
Indices of Different Regions in \pwn\label{table:flux}}
\tablehead{Region &\colhead{\phantom{aaa}} & \multicolumn{8}{c}{Flux Density (mJy)}
& \colhead{\phantom{aaa}} & \multicolumn{2}{c}{Spectral Index\tablenotemark{a}}
\\\cline{3-10}
& & \colhead{36\,cm} & \colhead{20\,cm} & \multicolumn{2}{c}{16\,cm} &
\multicolumn{2}{c}{6\,cm} & \multicolumn{2}{c}{3\,cm}} 
\startdata
All (pulsar+PWN) & & $110\pm30$ & $90\pm20$ & 65&10 & 65&10 & 45&6 && $-0.3$&0.1 \\
Pulsar && \nodata & 3.5\tablenotemark{b} & 2.00&0.05\tablenotemark{c} &
1.0&0.1 & 0.6&0.1 && $-1.04$&0.04\\
Bubble && \nodata & \nodata & 28 & 2 & 26.5 & 1 & 23.5 & 1 && $-0.13$&0.06\\
Head (excl.\ pulsar) && \nodata & \nodata & 16 & 3 & 16 & 2 & 15 & 2 && 0.0&0.2\\
Tail && \nodata & \nodata & 9&2 & 8&2 & 6&2 && $-0.2$ &0.3\\
Bubble northeast rim && \nodata & \nodata & 5.8&0.3 & 5.5&0.2 & 5.0&0.2 &&
$-0.10$&0.05\\
Bubble south rim & & \nodata & \nodata & 6.5&0.2 & 5.4&0.2 & 4.6&0.2 &&
$-0.24$&0.04
\enddata
\tablenotetext{a}{Spectral index $\alpha$ is defined as $S_\nu\propto
\nu^\alpha$.}
\tablenotetext{b}{Pulsed emission measured with the Parkes radio telescope
\citep{jw06}.}
\tablenotetext{c}{Pulsed emission measured with the pulsar binning data.}
\end{deluxetable*}

\subsection{Polarimetry}
\begin{figure*}[!th]
\epsscale{1.0}
\plottwo{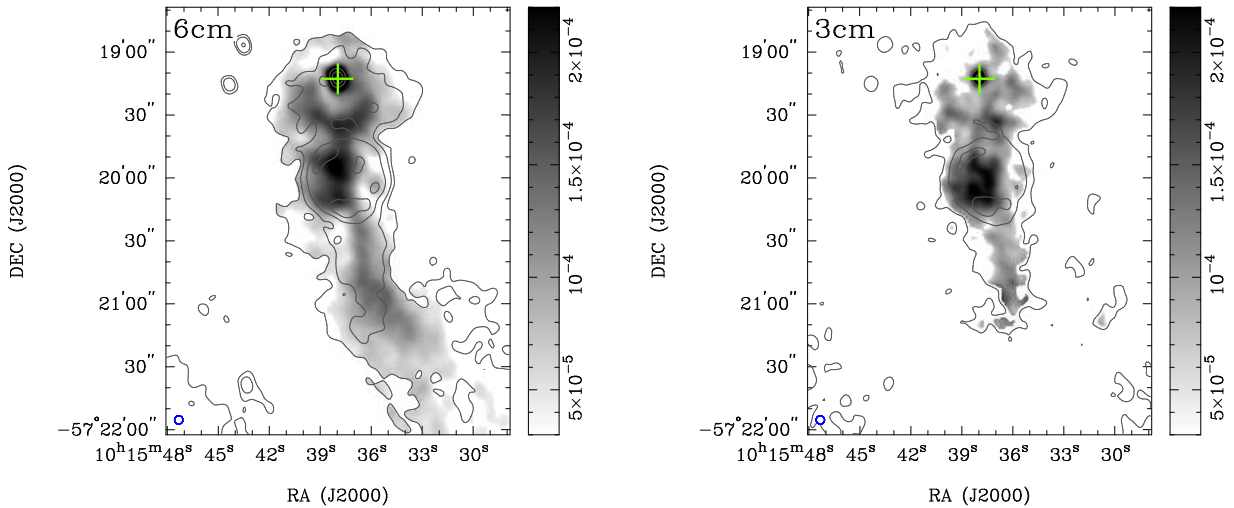}{3cm_pl.ps}
\caption{Linearly polarized intensity of \pwn\ at 3 and 6\,cm, overlaid with
total intensity contours at levels of 0.15, 0.2, 0.35, and 0.5\mjb. The maps are
clipped where the polarized intensity has S/N$<3$ or the total intensity has
S/N$<10$. The gray scales are linear, ranging from 0.03 to 0.22\mjb\ The
crosses mark the pulsar position and the beam sizes are shown in the lower
left. \label{fig:pl}}
\end{figure*}

Figure~\ref{fig:pl} shows the 6 and 3\,cm linearly polarized intensity maps. The
emission of \pwn\ is highly polarized, and the polarized flux generally
follows the total intensity, except it is enhanced near the bubble center and
rather faint in the west rim of the bubble. The fractional polarization of the
nebula is over 30\% in both bands. In particular, it is $\sim$50\% at the
bubble center and $\sim$30\% near the rim. The large-scale diffuse emission in
the southeast (see Figure~\ref{fig:field}) is unpolarized, hence it could be
background emission unrelated to the nebula. We do not show the 16\,cm result,
since only a very low degree of polarization was found. This can be attributed
to beam depolarization due to rapid variation in rotation measure (RM) across
the field (see below).

\begin{figure}[ht]
\epsscale{1.1}
\centering
\plotone{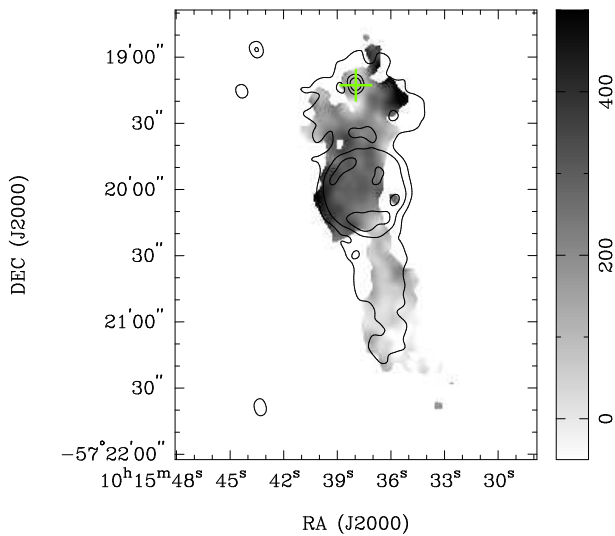}
\caption{RM distribution in \pwn\ obtained by comparing the 6 and
3\,cm polarization angles. The contours are from the 6\,cm total intensity
map at levels of 0.2, 0.35, and 0.5\mjb. The typical uncertainty is about
$50$\,rad\,m$^{-2}$ and the map is clipped where the uncertainty
$>80$\,rad\,m$^{-2}$. The gray scale is linear, ranging from $-50$ to
500\,rad\,m$^{-2}$. The cross marks the pulsar position. \label{fig:rm}}
\end{figure}

We estimated the foreground RM by comparing the polarization angles at
3 and 6\,cm, and the result is shown in Figure~\ref{fig:rm}. The RM changes
significantly across the nebula, from $\sim$100\,rad\,m$^{-2}$ in the head to
$\sim$300\,rad\,m$^{-2}$ near the bubble. It then drops rapidly to
0\,rad\,m$^{-2}$ in the tail. We found RM$\sim$100\,rad\,m$^{-2}$ near the
pulsar, consistent with the previously reported value of
$96\pm2$\,rad\,m$^{-2}$ \citep{jw06}. One potential issue with determining RM
from only two wavebands is the so-called n--$\pi$ ambiguity problem, which
arises when the polarization vectors rotate more than 180\arcdeg\ between the
two bands. We argue that this is not the case here, since otherwise it would
imply RM values over 1000\,rad\,m$^{-2}$, incompatible with the pulsar RM.
This is also supported by the lack of an abrupt jump in the RM map.

\begin{figure}[!ht]
\epsscale{1.2}
\centering
\plotone{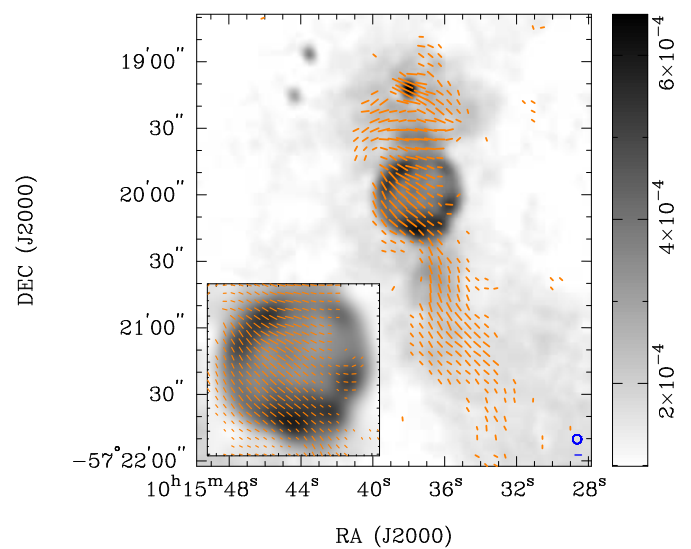}
\caption{The same 6\,cm intensity map of J1015 and \pwn\ as in
Figure~\ref{fig:wilkin}, overlaid with polarization $B$-vectors that indicate
the intrinsic magnetic field orientation. The vector lengths are proportional
to the polarized intensity at 6\,cm, with the scale bar in the lower right
representing of 0.1\mjb. Vectors with uncertainties $>10\arcdeg$ in PA are
clipped. The beam size is shown in the lower right. The inset is the same map
zoomed in at the bubble. \label{fig:bfield}}
\end{figure}

The RM map was then used to correct for the Faraday rotation of the
polarization vectors. The resulting map is shown in Figure~\ref{fig:bfield},
indicating the intrinsic orientation of the nebular magnetic field.
The field appears to wrap around the head in the north and runs orthogonal to
the overall nebular elongation in the interior of the head. Note that the
field orientation is somewhat different at the position of J1015 than in its
surroundings, which could be due to contamination by the pulsar emission.
Inside the bubble, the $B$-field changes direction for $\sim$45\arcdeg\ and
runs northeast--southwest, i.e.\ perpendicular to the northeast rim.
Intriguingly, the field follows closely the curvature of the southeast rim. It
is unclear if the west rim shows the same behavior, because the degree of
polarization there is too low. Just beyond the south rim, the field turns for
another $\sim$45\arcdeg\ and becomes parallel to the nebular elongation.
Further south, two similar switches of the field orientation are observed.
Altogether the changes in field direction seem quasi-periodic with a scale of
$\sim$30\arcsec.

Finally, we simulated a polarization observation at 16\,cm using the RM and
intrinsic polarization maps above. We found that, due to large RM values and
long wavelength, the polarization vectors rotate rapidly across the nebula.
The beam depolarization effect is therefore severe for the relatively low angular
resolution 16\,cm observation, resulting in a very low degree of polarization,
as observed.

\section{\bf Discussion}
\label{sec:discuss}
Our new radio observations reveal a remarkable nebula, \pwn\, surrounding J1015.
Its high degree of linear polarization and power-law spectrum indicate a
synchrotron nature for the emission.\footnote{Although thermal
bremsstrahlung radiation could also give a power-law spectrum, the observed
high degree of polarization excludes any significant contamination by this
kind of emission.} Fundamental questions are whether this nebula is physically
associated with the pulsar and what its nature is. We argue that this is a PWN
system powered by J1015 based on the following results: (1) the nebula is very
close in projection to the pulsar and, given its peculiar properties, a chance
coincidence seems very unlikely; (2) the head of the nebula and the pulsar
have comparable RM values, and the intrinsic $B$-field shows a continuous
pattern across the nebula; (3) the nebular morphology is broadly similar to
that of other PWNe, namely with a head that can be interpreted as a bow shock,
and a circular body like the one in the Guitar Nebula \citep{crl93}; and (4)
the large polarization fraction and a flat spectrum are both common
characteristics of radio PWNe
\citep[see][]{gs06}.

We should note that another possible (although less likely) interpretation of
the circular structure is a supernova remnant (SNR) with fresh electrons
provided by J1015. However, as we will show below, this implies an evolved and
highly sub-energetic SNR, and the ``tail'' of the nebula as part of the
remnant, which is not commonly observed. In the following discussion, we will
extract the physical parameters of the nebula according to these models.

\subsection{The ``Head'' of the Nebula as a Pulsar Bow Shock}
\label{sec:head}
The shape of the head is well fit by the analytic model of a bow shock
with an angular separation $\theta_0=20\arcsec$ between the pulsar and the
apex (see Figure~\ref{fig:wilkin}). For an inclined system, we show in
Appendix~\ref{sec:app_bs} that the projected shape can also be described by
the same model, but the true standoff distance is always smaller than
$\theta_0$, e.g., with inclination angles $i=60\arcdeg$, $45\arcdeg$,
and $20\arcdeg$ between the bow-shock axis and the line of sight (LOS), the
true standoff distance is smaller than the projected value by about 10\%,
20\%, and 50\%, respectively (see Figure~\ref{fig:proj_dist} in
Appendix~\ref{sec:app_bs}). In our case, this implies a standoff distance $r_0
\lesssim \theta_0 d\approx 0.5\,$pc at the pulsar distance of $d=5$\,kpc.
While the exact equality holds only for a side view of the bow shock (i.e.\
$i=90\arcdeg$), this should be a good estimate except when its axis is very
close to the LOS.

\begin{figure}[!ht]
\epsscale{1.0}
\centering
\plotone{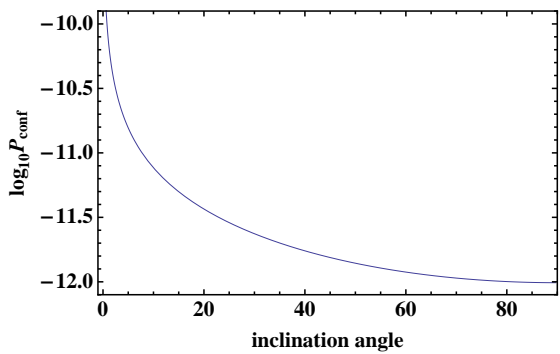}
\caption{Bow shock confinement pressure as a function of inclination angle.
 \label{fig:Pconf}}
\end{figure}
We can then estimate the confinement pressure of the pulsar wind to be
\begin{equation}
P_{\rm conf}=\frac{\dot E}{4\pi cr_0^2}\gtrsim 10^{-12}\,\mathrm{dyn\,cm}^{-2}
\end{equation} 
based on the pulsar spin-down luminosity $\dot E=8.3\times10^{35}$\ergs.
Its dependence on $i$ is plotted in Figure~\ref{fig:Pconf}. For
$i\sim90\arcdeg$, $P_{\rm conf}$ is comparable to the typical ISM pressure in
the Galactic plane \citep{fer01}, suggesting that the ram pressure is
relatively unimportant and hence the pulsar is moving at a low Mach number.
Only if the pulsar moves toward or away from us (i.e.\ small values of $i$),
could the confinement pressure be sensibly higher. Indeed, the head shows a
very diffuse morphology without a sharp boundary, which could indicate a weak
confinement and support the low Mach number scenario.

\subsection{Magnetic Field of \pwn}
\label{sec:bfield}
Since \pwn\ is not detected in other wavebands, we estimate the average
magnetic field strength based on the radio luminosity. By minimizing the total
energy of the particles and the field, we obtained the so-called equipartition
field 
\begin{equation}
B_{\rm eq}=(6\pi c_{12}L/V)^{2/7}, \label{eq:equib}
\end{equation}
where $L$ is the synchrotron luminosity, $V$ is the emission volume, and
$c_{12}$ is a constant depending on the spectral index \citep{pac70}. We
adopted the standard frequency range of $10^7$--$10^{13}$\,Hz and considered
only leptons in the wind with a filling factor of order unity. Taking the
bubble as a sphere and the head as a cone in 3D, we estimated $B_{\rm eq}\sim
70\,\mu$G and $\lesssim 60\,\mu$G\footnote{This is an upper limit assuming
$i$=90\arcdeg. For smaller inclination, the volume of the cone is larger.},
respectively. These values are similar to those found in other systems
\citep[e.g.,][]{ngc+10,nbg+12}. Their associated total energies are both about
$E_{\rm eq}\sim 10^{46}\,$erg. Note that any deviations from equipartition
would imply higher total energies.

If these structures are efficiently powered by the spin-down of J1015, the
time scale to fill them could be very short $\sim (E_{\rm eq}^{\rm
bubble}+E_{\rm eq}^{\rm head})/\dot E\sim 700$\,yr. The synchrotron cooling
time, on the other hand, is much longer. Even for particles emitting at the
highest observed frequency of $\nu\approx10$\,GHz (Figure~\ref{fig:spec}), the
cooling time
\begin{equation}
\tau_{\rm syn}=45\left(\frac{B}{\rm 10\,\mu G}\right)^{-3/2}
\left(\frac{\nu}{\rm 10\,GHz}\right)^{-1/2}\,\mbox{Myr}
\end{equation}
exceeds the pulsar's characteristic age of $\tau_c=39$\,kyr by three orders of
magnitude. It is therefore justified to neglect synchrotron loss and the radio
morphology of \pwn\ traces the motion of J1015, i.e.\ moving north at a PA of
$\sim$5\arcdeg. We can then infer another time scale from the offset $r_{\rm
offset}$ between the pulsar's current location and the bubble center (assuming
the bubble is at rest)
\begin{equation}
\tau_{\rm offs}=\frac{r_{\rm offset}}{v_{\rm psr,\perp}}\simeq1.2\times10^4\left(
\frac{v_{\rm psr,\perp}}{\rm 100\,km\,s^{-1}}\right)^{-1}\,\mbox{yr}.
\end{equation}
The pulsar projected velocity $v_{\rm psr,\perp}$ has not yet been measured,
and the only inference we can make about it from all the clues is a low Mach
number. If this is the case, $\tau_{\rm offset}$ could easily be of the order
of $\tau_c$. Incidentally, this coincidence seems to suggest that the bubble
started forming when the pulsar was much younger than now, which opens the
way to an alternative scenario, namely that the bubble coincides with the
position of an associated SNR. This scenario will be considered in
Section~\ref{sec:snr} below.

Interestingly, the projected spin axis of J1015 has a PA of 62\arcdeg\ (or
152\arcdeg\ if the pulsar emission is in the orthogonal mode) \citep{jw06}.
Either case shows a significant misalignment with the pulsar motion direction
(PA $\sim$5\arcdeg) inferred from the nebular elongation. While this may seem
uncommon among young isolated neutron stars \citep[see e.g.,][]{jhv+05,
nkc+12}, simulations of neutrino kicks indicate that a large misalignment
angle is more likely to be found in low-velocity pulsars \citep{nr07}. This
provides an indirect support for the low Mach number we argued above.

Our radio polarization measurement revealed a highly ordered magnetic field
structure in \pwn. As shown in Figure~\ref{fig:bfield}, the intrinsic
$B$-field wraps around the northern edge of the nebula. This is the first time
such a structure has been observed in bow shocks. It could indicate that the
magnetic field follows the shocked pulsar wind to sweep back due to the pulsar
motion. For bow shocks with high Mach number, the opening angle of the head is
generally much smaller. This structure is therefore more difficult to be
resolved by the observations. Inside the head, the $B$-field exhibits an
azimuthal geometry perpendicular to the nebular elongation, implying a helical
field trailing the pulsar. Unlike the more common case of a parallel
configuration, e.g., the Mouse and the Frying Pan \citep{yg05,nbg+12}, a
helical field has only previously been found in one bow shock PWN, namely,
G319.9$-$0.7 \citep{ngc+10}. The latter is believed to be moving faster than
J1015, at a few hundred \kms\ \citep{kkr+16}. While an azimuthal field could
be generated by a supersonically moving pulsar with aligned spin and velocity
\citep{rcl05}, this may not apply to J1015 as we have evidence for a large
misalignment angle. Our results imply that the $B$-field configuration in bow
shocks may depend more on the flow condition and other physical parameters
rather than the Mach number or the pulsar spin--velocity alignment.
Unfortunately the flow speed in \pwn\ is not easy to determine since no
synchrotron loss is observed in the radio band. It is essential to expand the
bow shock sample for further investigation.

The magnetic field changes direction quasi-periodically south of the bubble
and along the tail. Similar behavior has been found in two other bow shock
PWNe: the Mouse and G319.9$-$0.7 \citep{yb87,ngc+10}. Intriguingly, the latter
also exhibits a bulge morphology and the intrinsic field orientation switches
from azimuthal to parallel configuration near the bulge, as in our case. The
switches could be related to instabilities in the flow, which may have created
the bubble and the bulge, although the exact mechanism is still unclear. We
note that instabilities generally lead to a turbulent environment and it
remains to be explained how the high observed degree of polarization is
preserved. One possible solution is a large characteristic scale for the
turbulence, similar to what was found in the Snail \citep{mnb+16}. Further
studies with magnetohydrodynamic simulations are needed to understand the
cause and scale of the instabilities. Finally, turbulence in the interstellar
environment could also cause kinks in the tail and change the $B$-field
direction \citep[e.g., the Frying Pan;][]{nbg+12}, but it cannot produce the
bubble structure. 

\subsection{The Bubble as an Expanding Shell} \label{sec:bubble}
The equipartition $B$-field strengths in Section~\ref{sec:bfield} imply total
(magnetic + particle) pressures of at least $\sim200\times
10^{-12}$\,dyn\,cm$^{-2}$ and $\sim300\times 10^{-12}$\,dyn\,cm$^{-2}$ inside
the head and the bubble respectively, both much larger than the typical ISM
pressure of $\sim10^{-12}$\,dyn\,cm$^{-2}$. While in the head this could still
be consistent with a low pulsar Mach number (around 8--10), the overpressure
issue in the bubble is very significant (and could be even more severe in the
case of deviations from equipartition). Indeed as Figure~\ref{fig:bfield}
shows, the $B$-field generally follows the curvature of the
bubble rim in the southeast, indicating that it could be swept up by an
expanding shell \citep[see e.g.,][]{kb09,bp16}. This would compress the field
at the rim and result in higher emissivity than in the interior. Together
with projection effects, these lead to the limb-brightened structure of the
bubble. The surface brightness of the bubble peaks in the northeast and the
southwest where the rim is perpendicular to the $B$-field. Similar behavior is
typically observed in young SNRs \citep[e.g.,][]{dg15} and the enhanced radio
emission is interpreted as evidence of efficient particle acceleration when
the magnetic field is quasi-parallel to the shock direction. However this does
not work in our case, since it would induce a strongly turbulent field and
re-acceleration of the pulsar wind, resulting in a very low degree of
polarization and a steep radio spectrum incompatible with the observations.

We instead propose another scenario: that the bubble acts as a ``magnetic
bottle'' and traps the relativistic particles. Due to conservation of magnetic
flux, the expansion of the bubble creates a gradient in the magnetic field
along the northeast--southwest direction. A particle moving toward the
equatorial zone would then show a decrease in the pitch angle and an increase
in the longitudinal velocity. The opposite occurs when it moves away from the
equatorial zone. Particles entering from one side of the bubble should, in
principle, be able to exit from the other side. However, the presence of
scattering broadens the pitch angle distribution. Some particles could
therefore have pitch angles over 90\arcdeg\ before exiting the bubble, i.e.\
they would be mirrored back. The lower polarization fraction observed near the
rim of the bubble could suggest a mild level of turbulence, providing some
degree of scattering and helping the confinement.

Note that in order for the electrons to bounce back and forth, the mean free path
parallel to the direction of the ordered magnetic field ($\lambda_\parallel$)
should be no smaller than the bubble radius ($R_{\rm bubble}$). 
The former is proportional to the gyroradius ($r_g$) with
\begin{equation}
\lambda_\parallel=\eta r_g\sim3.5\times 10^{12}\left(\frac{\nu}
{\rm 5\,GHz}\right)\left(\frac{B}{\rm 10\,\mu G}\right)^{-3/2}\eta\,\mbox{cm},
\end{equation}
where $\eta=(\delta B/B)_{\rm res}^{-2}$ is 
related to the perturbation of the magnetic field resonant with the radio
emitting particles \citep{rey98}.
The condition $\lambda_\parallel \gtrsim R_{\rm bubble}$ implies 
\begin{equation}
\left(\frac{\delta B}{B}\right)_{\rm res}\lesssim 1.5\times10^{-3}\left(\frac{\nu}
{\rm 5\,GHz}\right)^{1/4}\left(\frac{B}{\rm 10\,\mu G}\right)^{-3/4}.
\end{equation}
This in not unreasonable, considering that in a (Kolmogorov-like) turbulent
cascade $(\delta B / B)_{\rm res}$ decreases at smaller scales. This suggests
$\eta\gg1$, hence diffusion perpendicular to the magnetic field has a small
mean free path ($r_\perp\approx r_g/\eta$) and is very inefficient.
As a result, electrons can be considered as frozen into their original flux tubes.

The synchrotron emission of the bubble primarily comes from particles injected
by the pulsar wind with the magnetic field mostly swept-up from the surroundings.
The observed limb brightening can be explained by:
(1) particles near the center emit mostly in directions off the observer LOS,
due to the geometry of the magnetic bottle,
(2) they also emit less effectively due to generally lower pitch angles, and
(3) the higher longitudinal velocities of the particles in the
equatorial zone imply a lower number density and hence a lower emissivity
compared with the polar regions.

From a dynamical point of view, these trapped particles would be responsible
for most of the pressure inflating the bubble. It can be shown
that, for a Kolmogorov-like power spectrum of magnetic fluctuations, the
parallel mean free path of particles is proportional to their Lorentz factor
as $\gamma^{-2/3}$; i.e.\ higher-energy electrons (than the radio-emitting
ones), which carry most of the energy, reach an isotropic velocity
distribution more efficiently inside the bubble, likely providing the
required pressure. The expansion of the bubble can only be balanced by the
inertia of the ambient medium. A simple dimensional argument suggests that in
this case the total
energy of the bubble must be comparable with $\sim$$\rho_0 R_{\rm
bubble}^5/t_{\rm exp}^2$, where $\rho_0$ is the ambient density and $t_{\rm
exp}$ is the bubble expansion time. We can then use the equipartition
$B$-field result in Equation~(\ref{eq:equib}) to estimate the ambient density
\begin{equation}
n_0\sim 10^3\left(\frac{t_{\rm exp}}{\tau_c}\right)^{2}\,\mbox{cm}^{-3}.
\end{equation}
If $t_{\rm exp}\approx \tau_{\rm offs}$ is of the order of $\tau_c$, the
environment must be very dense to keep the bubble confined. Otherwise a very
young bubble would imply a high pulsar velocity to explain the observed
offset, and this is in contrast to what was previously deduced on the basis of
the properties of the nebular head. It is therefore more likely that the
bubble is expanding into a dense medium.

Driven by the above considerations, we have then checked whether there is
evidence of a dense surrounding medium. In the optical catalogs
\citep{fs84,db02}, we found two dark clouds, FeSt2-78 and FeSt2-79, coinciding
with \pwn\ (see Figure~\ref{fig:field}). This LOS is near the
Sagittarius--Carina arm tangent ($l=282\arcdeg$), which has a distance of
$\sim$4\,kpc \citep{gra70}, broadly consistent with that of J1015. Hence, the
clouds are possibly in the pulsar vicinity. Based on the difference between
the central and background reddening of the clouds \citep{db02}, we estimate
their optical extinction and hydrogen column densities using empirical
relations \citep{fit04,go09}. The latter give densities of $n_0\sim
450$\,cm$^{-3}$ and $\sim$680\,cm$^{-3}$ for FeSt2-78 and FeSt2-79,
respectively, assuming spherical clouds with 6\arcmin\ angular diameter
\citep{db02} at the pulsar distance of 5\,kpc. Following the same procedure,
we estimate the density of all dark clouds in the above catalogs with $\Delta
l=\pm 20\arcdeg$ from J1015 and $|b|<1\arcdeg$. There are only five objects
with density higher than 450\,cm$^{-3}$ and their total area in the sky
suggests a chance coincidence of 0.1\% to have a dense cloud overlapping with
\pwn. We argue that the PWN is in a dense environment with $n_0$ of a few
hundred cm$^{-3}$, as inferred from our analysis.

\subsection{Origin of the Bubble}
We now turn to the origin of the bubble. The flat radio spectrum of the bubble
indicates that it is part of the PWN structure. There are several possible
scenarios in which an extended nebula could be formed at a distance from the
pulsar:
(1) a relic nebula crushed by the supernova reverse shock,
(2) injection of pulsar wind into a pre-existing cavity such as a stellar wind
bubble,
(3) sudden expansion of the flow due to mass loading \citep{mlv15}, and
(4) an expanding bubble driven by flow instabilities \citep[e.g.,][]{vi08}.
The first two scenarios would be viable only if the associated shell-type SNR
no longer emits in radio and the ordered longitudinal magnetic field between
the pulsar and the bubble could be explained. The $B$-field and the limb
brightness structure of the bubble do not seem to match what is seen in a
relic PWN, such as G327.1$-$1.1 \citep{tsk+15,mnb+16}. Also in this case the
``tail'' of the nebula would be a blowout of the SNR and happen to be in the
opposite direction to the pulsar by chance. Finally, mass loading generally
does not produce a circular structure.

\begin{figure*}[th]
\epsscale{0.8}
\centering
\plottwo{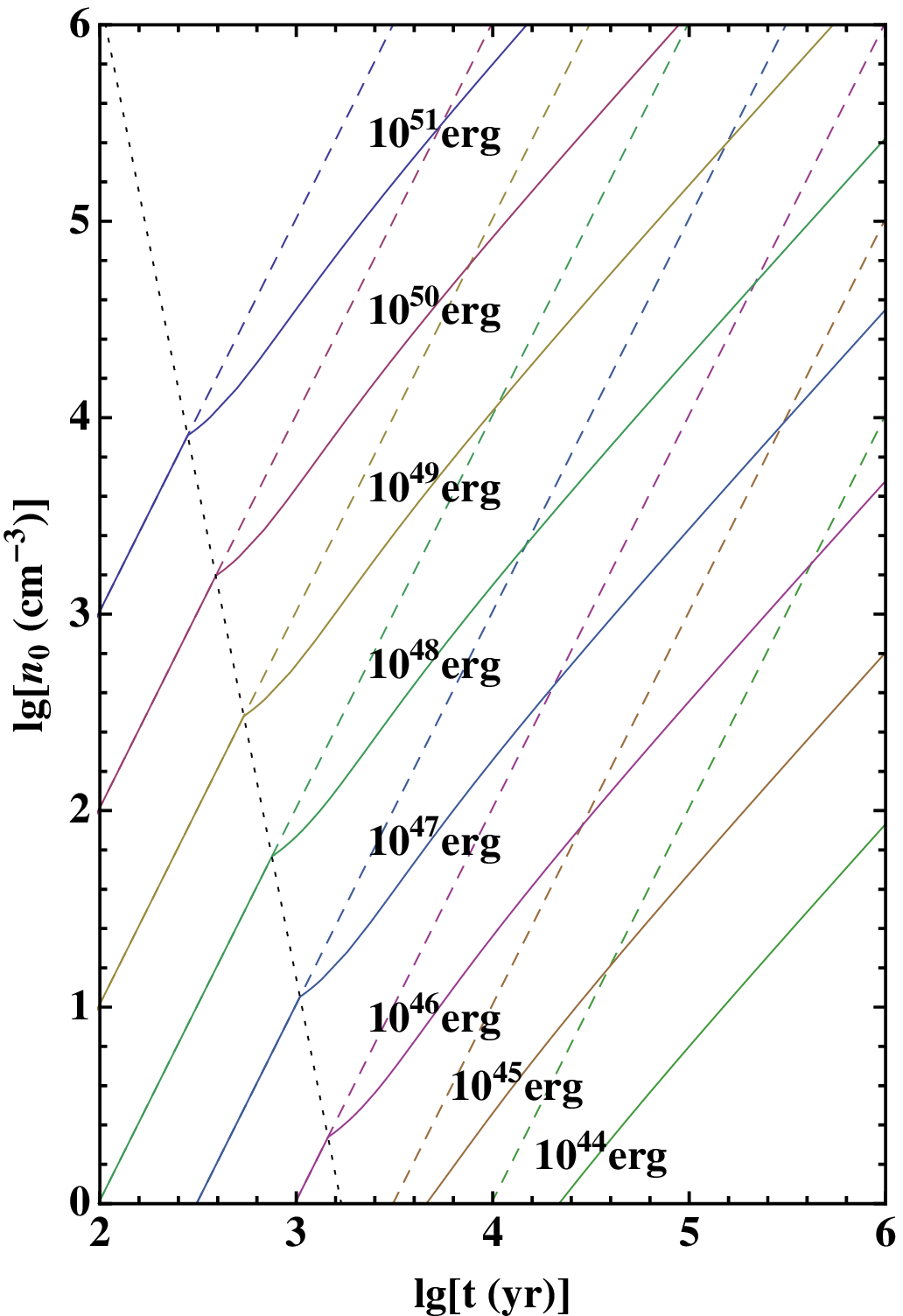}{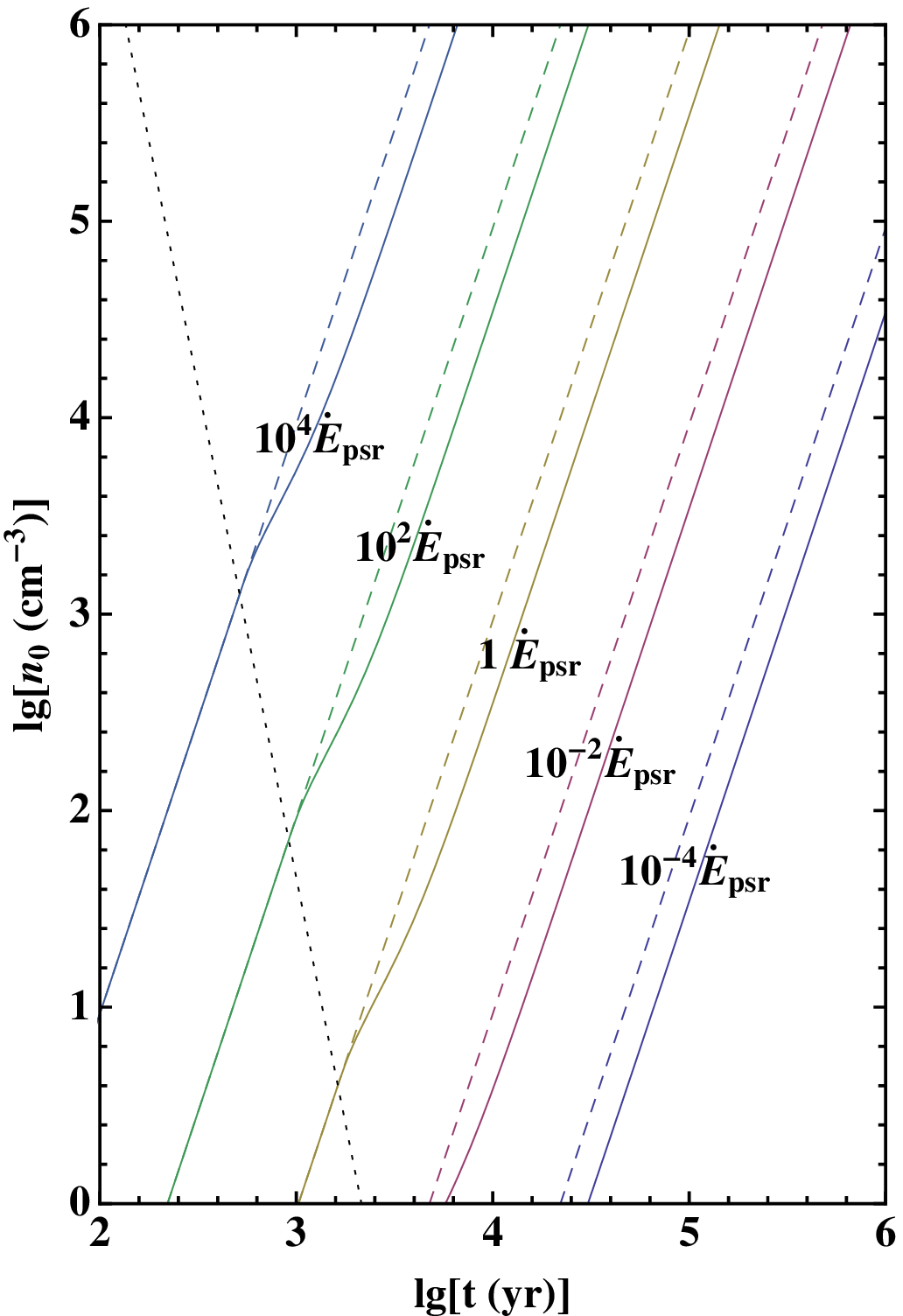}
\caption{Bubble expansion time ($t$) and ambient density ($n_0$) that 
give the observed bubble radius of 0.5\,pc. Left: the case of sudden
initial energy release as in an SNR, according to the evolution model by
\citet{bp04}. The solid color lines represent different total energies. The
transition from the Sedov phase to the radiative phase, which is indicated by
the black dotted line, leads to a power-law evolution with a different slope.
The dashed lines show the extrapolation of the adiabatic evolution. It is
apparent that using the adiabatic extrapolation largely overestimates $n_0$ for
a given $t$. Right: the same for a bubble with constant energy
injection rates. The evolution curves transit from adiabatic to radiative
regimes smoothly, and eventually approach a trend parallel to the original
one. \label{fig:density_age}}
\end{figure*}

To explore the instabilities scenario, we follow \citet{vi08} to assume that
the post-shock flow became unstable downstream, injecting a fraction $f_{\dot
E}$ of the spin-down energy into the bubble. The total injected
energy into the bubble is
\begin{equation}
E_0\simeq f_{\dot E}\dot E t_{\rm exp}\simeq10^{48}f_{\dot
E}\left(\frac{t_{\rm exp}}{\tau_c}\right)\,\mbox{erg}.
\end{equation}
If the flow instability occurred in a short timescale and the bulk of the
energy was released a long time ago, the evolution should be similar to that
of an SNR, in spite of the different energy scale. We employed an analytic
model developed by \citet{bp04} to follow both the adiabatic and radiative
phases of the SNR evolution \citep{tm99}. The bubble radius in the adiabatic
phase can be described by the standard Sedov--Taylor solution,
\begin{equation}
R_{\rm bubble}(t_{\rm exp})=1.152
\left(\frac{\edot}{\rho_0}\right)^{1/5}t_{\rm exp}^{2/5}.
\end{equation}
The transition to the radiative phase occurs at time and radius of
\begin{equation}
t_{\rm tran}\simeq 2.9\times10^4\left(\frac{E_0}{\rm 10^{51}\,erg}
\right)^{4/17}\left(\frac{n_0}{\rm 1\,cm^{-3}}\right)^{-9/17}\mbox{yr}\quad\mbox{and}
\end{equation}
\begin{equation}
R_{\rm tran}\simeq 19.1\left(\frac{E_0}{\rm 10^{51}\,erg}\right)^{5/17}
\left(\frac{n_0}{\rm 1\,cm^{-3}}\right)^{-7/17}\mbox{pc},
\end{equation}
and the radiative phase can be described by the inverse evolutionary relation
\begin{equation}
t'(R')=\frac{2}{35}\sqrt{R'-1}\left(5R'^3+6R'^2+8R'+16\right)-0.248,
\end{equation}
where $t'=t/(1.14\,t_{\rm tran})$ and $R'=R/(0.85\,R_{\rm tran})$.
Figure~\ref{fig:density_age} (left) shows a diagnostic plot of the ambient
density and age that gives the observed $R_{\rm bubble}=0.5$\,pc. For
$n_0\sim500$\,cm$^{-3}$ and $t_{\rm exp}\sim4\times10^4$\,yr, a total energy
of $\sim10^{47}$\,erg is sufficient to give rise to the bubble, i.e.\ $f_{\dot
E}$ of the order of 10\%. We also illustrate in the plot estimates with only
the Sedov--Taylor solution: in this case the constraints on $n_0$ are
overestimated by an order of magnitude for a given age. This highlights the
importance of correctly accounting for the transition to the radiative phase.

If instead the bubble has been constantly powered by J1015 with constant
\edot\ until the present time, \citet{dok02} developed the adiabatic solution  
\begin{equation}
R_{\rm bubble}(t_{\rm exp})=0.929
\left(\frac{\edot}{\rho_0}\right)^{1/5}t_{\rm exp}^{3/5}.
\end{equation}
We show in Appendix~\ref{sec:app_snr} that this solution also applies to the
asymptotic radiative phase with a
scaling factor of 0.821. The result is plotted in Figure~\ref{fig:density_age}
(right). For $n_0\sim500$\,cm$^{-3}$ and $t_{\rm exp}\sim4\times10^4$\,yr,
$f_{\dot E}\sim1\%$ is sufficient.

\subsection{Where is the SNR?}
\label{sec:snr}
With a characteristic age of only 39\,kyr, the true age of J1015 should
be of the same order or smaller, such that the parent SNR may still be
visible. As discussed, the elongation of \pwn\ indicates the
pulsar's direction of motion. In this region of the sky there are only SNR
G284.3$-$1.8 \citep{wrk+15} and a candidate SNR G282.8$-$1.2 \citep{mcg02}.
However, neither of them aligns with the nebula. There is also no known
massive star cluster in the general direction. As we argued in
Section~\ref{sec:head}, the pulsar could be traveling at a low velocity,
hence the supernova site should not be too far from the pulsar's current
location, possibly within the same dense clouds we found in the catalog. If
this is the case, the SNR could have already evolved to a late stage and
become unobservable.

Another possibility to investigate is whether the radio bubble is indeed 
the parent SNR of J1015. If the pulsar were born at the bubble center, the
offset to its current location would imply a traverse velocity
significantly larger than the SNR expansion rate, with $v_{\rm
psr,\perp}/v_{\rm snr}=r_{\rm offset}/R_{\rm bubble}=2.5$. A more serious
issue with this picture is that the high ambient density observed is not
sufficient by itself to account for the small physical size of the bubble if
it is an SNR; an exceptionally sub-energetic supernova explosion is also
required. As we discussed in Section~\ref{sec:bubble} and showed in
Figure~\ref{fig:density_age}, the observed bubble size implies a total
energy of $\sim10^{47}$\,erg. This is much lower than the typical supernova
energy of $10^{51}$\,erg.

In this scenario, the sub-energetic SNR has evolved well beyond the Sedov
phase, thus the particle acceleration efficiency should have decreased
considerably \citep{bp10}. Although the magnetic structure of \pwn\ belongs to
the relic SNR, it is only detectable in radio because leptons are provided by
the pulsar to power the synchrotron emission. This is also what the flat radio
spectrum seems to suggest. This situation could be similar to the case of SNR
G5.4$-$1.2, which shows a pulsar and associated PWN outside (while not far
from) the SNR boundary. The radio spectral index of the SNR limb is flatter
when close to the pulsar position \citep{fkw94}, a fact that could be
attributed to the radio emission from leptons with a hard energy distribution
leaking from the PWN. In short, the interpretation of the bubble as an SNR
would be tenable under various aspects, except for the requirement of $E_0$
about $10^4$ times smaller than the average, making it rather unlikely.

\section{\bf Conclusion}
\label{sec:concl}
We presented radio observations of the field of \psr\ at 36, 16, 6, and 3\,cm
taken with MOST and ATCA. The radio images revealed a nebula, \pwn, consisting
of a diffuse head, a circular bubble, and a collimated tail. Based on its
positional coincidence with the pulsar, its flat spectrum, and high degree of
linear polarization, we suggest that the source is a newly discovered PWN
system with the head as a bow shock moving at a low Mach number and the bubble
as a shell expanding in a dense medium. We also considered an alternative
scenario with the bubble as the parent SNR of the pulsar. However, this
requires a supernova explosion energy much smaller than the canonical value,
which does not seem plausible.

\pwn\ presents a rare example of a slow-moving bow-shock PWN, in which the
pulsar spin axis misaligns with the proper motion direction. On the other
hand, it shows similar $B$-field geometry and expanding bubble feature as in
other higher Mach number systems. This implies that the bow shock properties
could depend more on other factors, such as the flow condition, instead of the
pulsar velocity.

\begin{figure*}[!th]
\centering
\includegraphics[width=0.45\textwidth, bb=18 185 594 600]{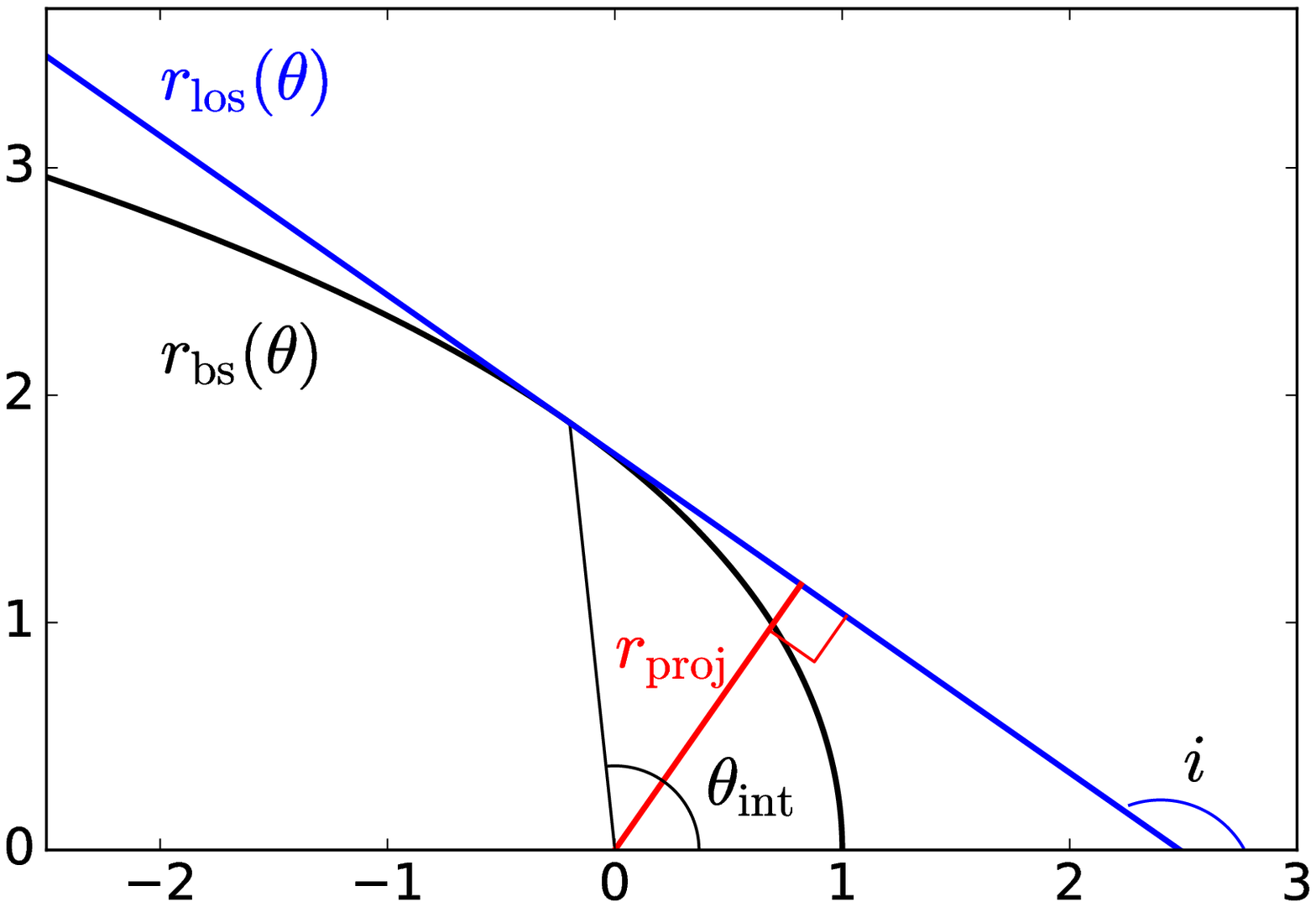}
\includegraphics[width=0.4\textwidth]{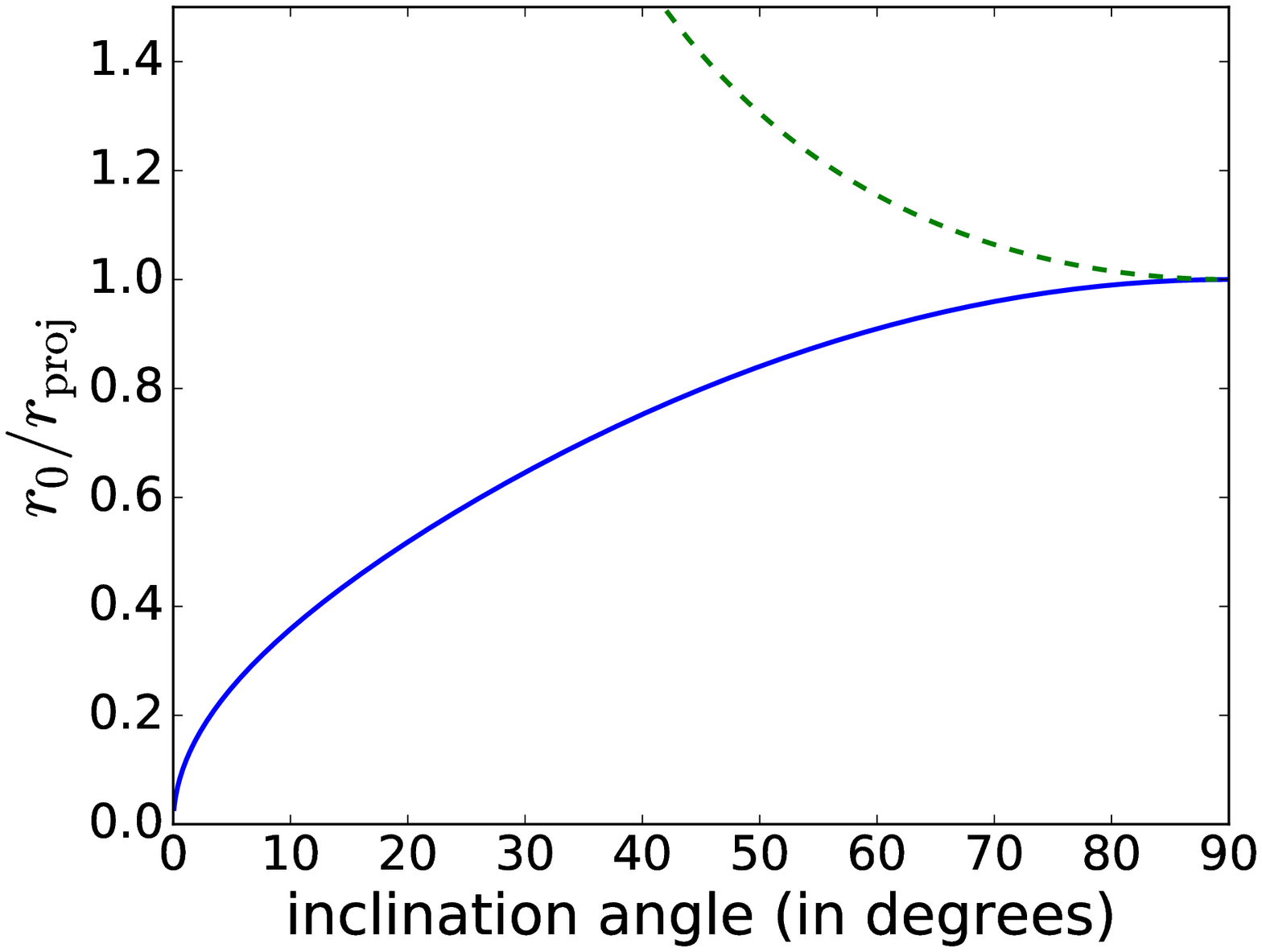}
\caption{Left: illustration of the projected bow-shock standoff
distance $\rproj$. The black line is the bow-shock model given by
\citet{wil96} and the blue line shows the line of sight with an inclination angle
$i$. The axes are in units of $r_0$, the true standoff distance. Right:
the ratio between $r_0$ and $\rproj$ as a function of inclination angle
(blue solid line). Note that the $\rproj$ is always larger than $r_0$ and the
result is invariant under a transform of $i$ to $180\arcdeg -i$. A simple
$1/\sin i$ relation is plotted by the green dashed line for comparison.
\label{fig:proj_dist}}
\end{figure*}

\acknowledgements
We thank the referee for careful reading and useful suggestions, Zhongxiang
Wang for providing the IR and optical data, and C. Woo for assistance with
some of the graphical works for this paper. The Molonglo Observatory site
manager, Duncan Campbell-Wilson, was responsible for the smooth and efficient
operation of the telescope at a time when sensitivity was seriously
compromised by strong RFI. The MOST is owned and operated by the University of
Sydney, with support from the Australian Research Council and the School of
Physics. The Australia Telescope is funded by the Commonwealth of Australia
for operation as a National Facility managed by CSIRO. C.-Y.N.\ is supported
by an ECS grant of the Hong Kong Government under HKU 709713P.

{\it Facilities: ATCA, Molonglo Observatory.}
{\it Software: MIRIAD \citep{stw95}.}

{\bf \appendix }
\section{\bf Projected Standoff Distance and Shape of an Inclined Bow Shock}
\label{sec:app_bs}
\setcounter{equation}{13}

We start from the analytic formula derived by \citet{wil96}, which describes a
bow shock shape in polar coordinates:
\begin{equation}
\rbs(\tht)=r_0|\csc\tht|\sqrt{3(1-\tht\cot\tht)},
\label{eq:Rbs}
\end{equation}
where $\rbs$ is the radial distance from the origin (i.e.\ the position of the
star) and $-\pi<\tht<\pi$ is the polar angle from the direction of the
standoff point, expressed in radians. The shape described by the above
equation is universal, while the relative strength of the stellar wind and the
ambient ram pressure determine only the spatial scale of the bow shock, namely
$r_0$. Of course, the exact shape described by Equation~(\ref{eq:Rbs}) could
actually be observed only when the viewing angle is orthogonal to the stellar
velocity direction (i.e.\ inclination angle $i=90^\circ$). We investigate here
how the projected standoff distance, $\rproj$, as well as the projected shape
of an inclined bow shock, depend on the inclination angle.

We first limit our analysis to the symmetry plane identified by the direction
of the standoff point and the LOS and passing through the
origin. The LOS with slope $b=\tan i$ ($i$ is defined modulo $\pi$ radians)
and a generic projected displacement $\aproj$ from the origin can be described
by
\begin{equation}
\rlos(\tht)=\frac{\aproj\sqrt{1+b^2}}{b\cos\tht-\sin\tht}
\label{eq:Rlos}
\end{equation}
in polar coordinates. Note that $\rlos$ changes sign if $\tht$ shifts by
$\pi$ radians. However, in polar coordinates it identifies the same point. For
a given value of $b$, if the bow shock and the LOS
intersect at $\rbs(\thi)=\rlos(\thi)$, the projected distance from the origin
is then
\begin{equation}
\aproj(\thi)=r_0\,{\rm sgn}(\thi)(b\cot\thi-1)\sqrt{\frac{3(1-\thi\cot\thi)}{1+b^2}}.
\label{eq:ap}
\end{equation} 
Consider a family of LOS with the same slope $b$ but intersecting the bow
shock at different $\thi$; the one tangent to the bow shock should have
largest $\aproj$ and hence $\aproj=\rproj$ in this case (see
Figure~\ref{fig:proj_dist} left). This can be done analytically by putting
${\rm d}\aproj(\thi)/{\rm d}\thi=0$. The resulting formula is rather
cumbersome, but it can be simplified to 
\begin{eqnarray}
(\sin\thi)(\thi-\sin\thi\cos\thi)-b\left[3\cos\thi\right. \nonumber\\
\times \left.(\thi-\sin\thi\cos\thi)-2\sin^3\thi\right]=0.
\end{eqnarray}
Solving this equation for $\thi$ is very complex, but the
solution for $b$ is trivial:
\begin{eqnarray}
b(&\thi&)=\tan[i(\thi)]\nonumber\\
&=&\frac{\sin\thi(\thi-\sin\thi\cos\thi)}{3\cos\thi(\thi-\sin\thi
\cos\thi)-2\sin^3\thi}
\label{eq:bsol}
\end{eqnarray}
(a sign change in $\thi$ implies a sign change in $b$ and hence in
$i$). Hence, for any values of intersection $\thi$, we can easily
derive the related inclination angle: for that angle, and only for that,
Equation~(\ref{eq:ap}) actually gives $\rproj$. Figure~\ref{fig:proj_dist} (right)
shows the resulting $\rproj$ as a function of $i$, obtained from a parametric
plot with $\thi$ as the variable and the $x$ and $y$ values from
Equations~(\ref{eq:bsol}) and (\ref{eq:ap}). The result indicates that $\rproj$ is
always larger than the true standoff distance $r_0$ for any $i$ (except for
$i=\pm90\arcdeg$, when it is equal to $r_0$) and it is different from the
simple $\sin i$ dependence. The latter, which was used by some authors in
previous studies to correct for the orientation, is therefore not even
qualitatively correct.

\begin{figure*}[th]
\epsscale{1.0}
\centering
\plotone{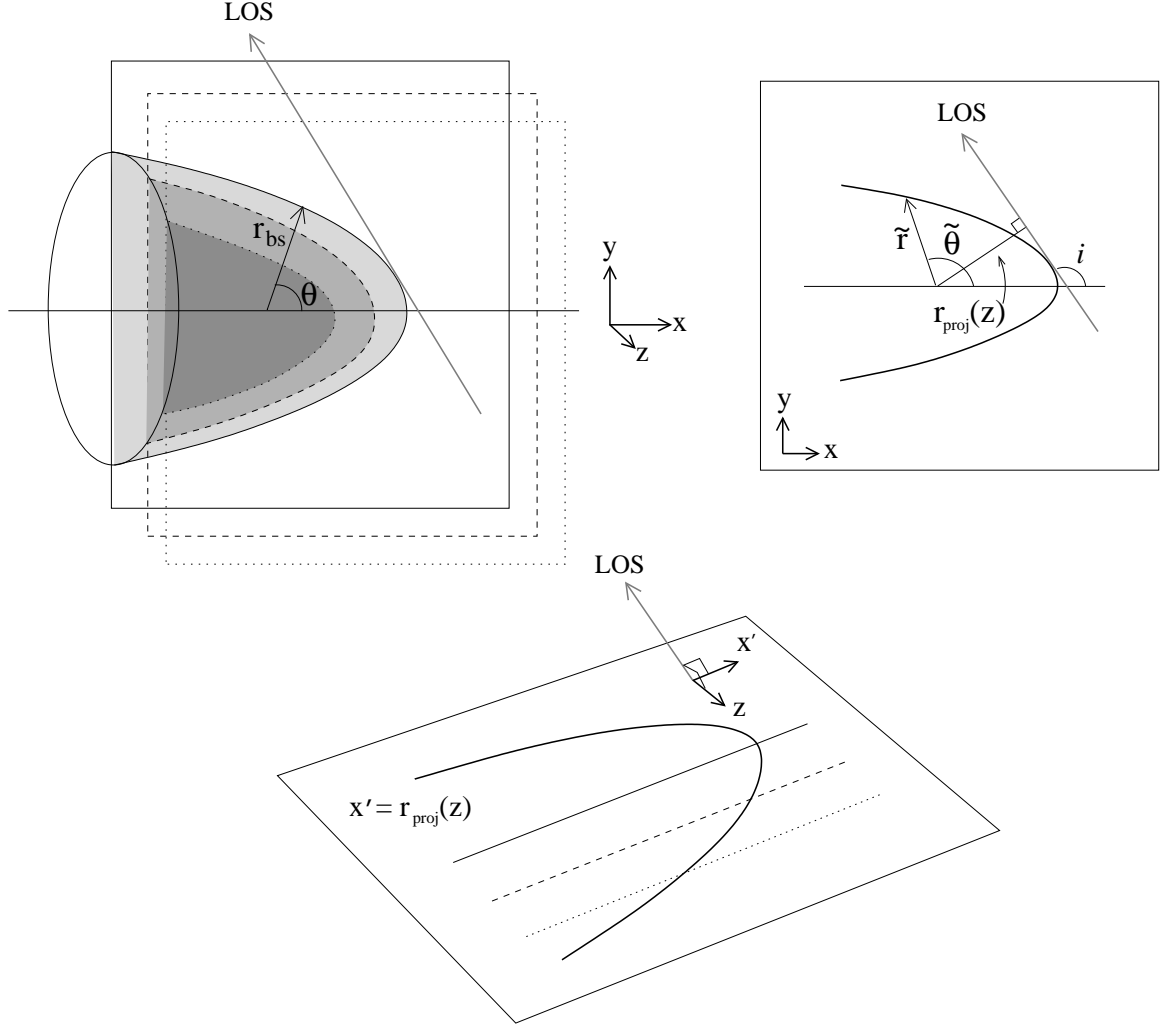}
\caption{Illustration of a bow shock projected onto the plane of the sky and
sliced in the $z$ direction. This also indicates the choice of coordinate system
and symbols used in the calculation.
\label{fig:bs_3d}}
\end{figure*}

A similar approach can also be used to derive the position of the projected
limb along any LOS, and ultimately, to derive the shape of the bow shock
projected at a generic inclination angle. We define a coordinate system such
that: $x$ is directed along the bow shock axis, toward its apex; $y$ is the
orthogonal axis lying on the plane defined by $\hat x$ and the LOS; and
$z$ is orthogonal to the previous two. Define $x'$ as the $x$-axis projected
onto the plane of the sky, then the shape of the projected bow shock is given
by a function $x'=\rproj(z)$, where the $z$ coordinate of a point gives the
displacement from the projected axis (see Figure~\ref{fig:bs_3d}). We first
slice the bow shock along the $z$ direction. For a given $z$ value, the slice
of the bow shock in the $x$--$y$ plane can be described by a function $\tr
(\ttht)$, where $\ttht$ is the 2D polar angle. $\tr$ is related to $\rbs$ and
$z$ by
\begin{equation}
\rbs^2 = \tr^2+z^2 \quad\mbox{and}
\end{equation}
\begin{equation}
\rbs\cos\theta = \tr\cos\ttht.\label{eq:rbs}
\end{equation}
In Equation~(\ref{eq:rbs}), the condition to have a solution for
$\ttht$ is
\begin{equation}
z^2\leqslant \rbs^2\sin^2\tht=3(1-\tht\cot\tht).\label{eq:th_min}
\end{equation}

\begin{figure*}[th]
\epsscale{1.0}
\centering
\plottwo{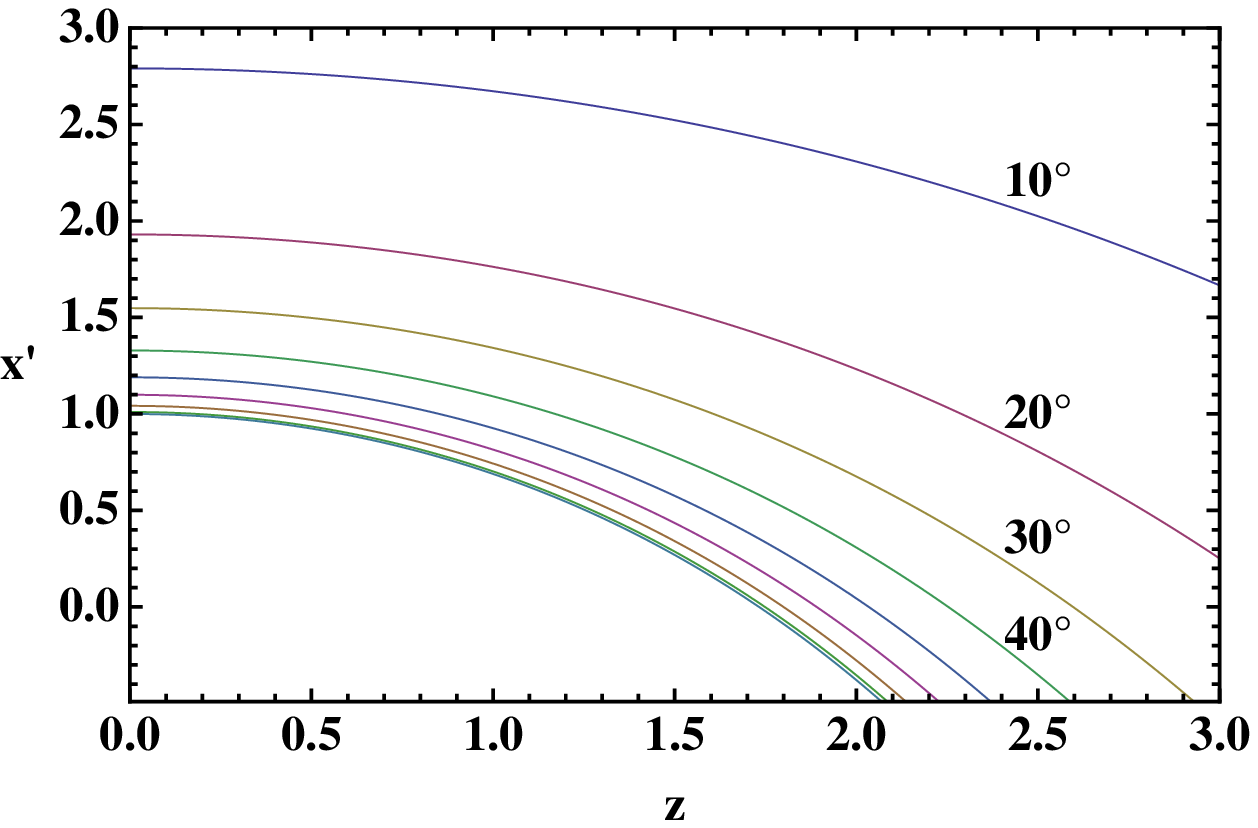}{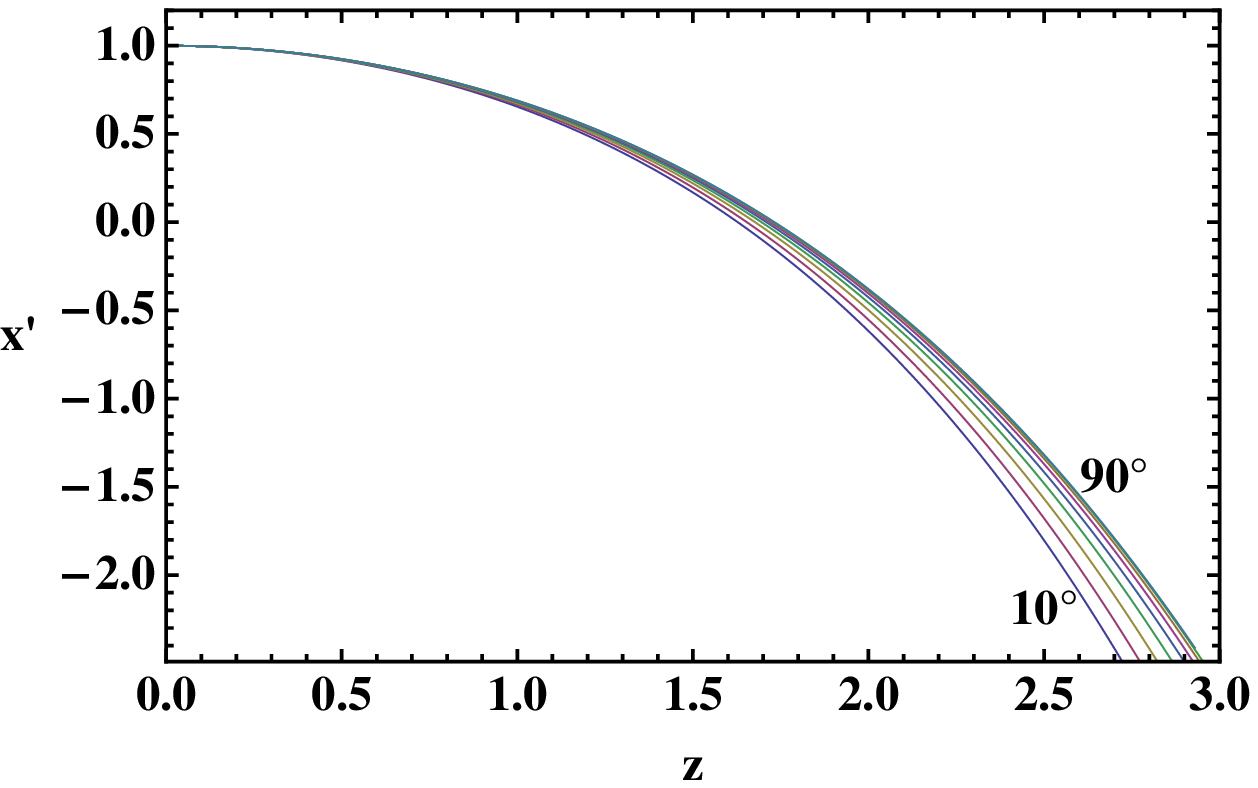}
\caption{Left: projected shapes of a bow-shock model by \citet{wil96}
with different inclination angles. Right: the same plot scaled by
$1/\rproj(\thi,0)$ to match the projected standoff distance. In both plots, the
axes are in units of $r_0$ and the lines represent different inclination
angles of from 10\arcdeg\ to 90\arcdeg. (See Figure~\ref{fig:proj_dist} for
the definition of the inclination angle, $r_0$, and $\rproj$.)
\label{fig:proj_shape}}
\end{figure*}

In an analogous way as we did with Equation~(\ref{eq:Rlos}), we define
\begin{equation}
\xlos(\ttht)=\frac{\aproj\sqrt{1+b^2}}{b\cos\ttht-\sin\ttht}
\label{eq:RlosX}
\end{equation}
and take its intersection with $\tr$ at $\tthti$, i.e.\
$\tr(\tthti)=\xlos(\tthti)$.
After some simplification, we obtain the projected elongation (the equivalent
in 3D of Equation~(\ref{eq:ap}))
\begin{eqnarray}
&\aproj(\thi,z)=r_0\left[b\cot\thi\sqrt{3(1-\thi\cot\thi)}\right.\nonumber\\
&-\left.\sqrt{3(1-\thi\cot\thi)-z^2/r_0^2}\right]\bigg/\sqrt{1+b^2}.
\label{eq:apX}
\end{eqnarray}
Following the same procedure as before, we take the derivative and finally
derive
\begin{eqnarray}
b(&\thi&,z)=\tan[i(\thi,z)]=\sqrt{\frac{3(1-\thi\cot\thi)}{3(1-\thi\cot\thi)-z^2/r_0^2}}
\nonumber\\
&\times&
 \frac{\sin\thi(\thi-\sin\thi\cos\thi)}{3\cos\thi(\thi-\sin\thi\cos\thi)-2\sin^3\thi}.
\label{eq:bsolX}
\end{eqnarray}
As a check, for $z=0$, the factor within the square root reduces to unity, and
hence Equation~\ref{eq:bsol} is recovered.

As illustrated in Figure~\ref{fig:bs_3d}, for a given offset $z$, figures
equivalent to Figure~\ref{fig:proj_dist} (right) can now be easily obtained as
parametric plots of the variable $\thi$ (where $\thi$ satisfies the condition
in Equation~(\ref{eq:th_min})) using Equations~(\ref{eq:bsolX}) and
(\ref{eq:apX}). This gives
$\rproj(\thi,z)$, the coordinate of the projected limb along the $x'$ axis. By
performing this calculation for all $z$ and $\thi$ values, $\rproj$ for all
values of $i$ can be obtained. On the other hand if $i$ is given, we need to
solve Equation~(\ref{eq:bsolX}) to get $\thi(z,i)$ (this is the only part of the
procedure that still needs to be done numerically). The projected profiles of
different inclination angles are shown in Figure~\ref{fig:proj_shape} (left). As
$i$ decreases from 90\arcdeg, $\rproj$ increases but the overall shape remains
similar. In real life, $\rproj$ is often the only observable while $i$ or
$r_0$ are unknown; we therefore scale the plot by $1/\rproj(\thi,0)$ for
comparison and the result is shown in Figure~\ref{fig:proj_shape} (right).
Obviously, the change in the projected shape is minimal and thus very
difficult to distinguish between them from observations. Hence, we conclude
that the same analytic solution is a sufficiently good approximation also for
inclined cases with $i$ different from 90\arcdeg.

\section{\bf Expansion of a Bubble with Continuous Injection in a Uniform Medium}
\label{sec:app_snr} We present here a thin-shell model of an expanding bubble
in a homogeneous ambient medium, in the case of a continuous energy release at
a constant rate \edot. The approach is similar to that presented by
\citet{bp04}, for an initial energy of the bubble. We will also model the
transition from adiabatic to radiative phases.

The basic assumptions are that matter is confined to a thin, spherically
symmetric shell at the boundary of the bubble, which contains all the kinetic
energy of the system, while all the thermal energy is contained in the
homogeneous interior of the bubble. The conservation of mass and momentum in
the shell are described by the equations
\begin{equation}
\frac{{\rm d}M}{{\rm d}t} = 4\pi\rho_0R^2\dot R\quad\mbox{and}	
\end{equation}
\begin{equation}
\frac{{\rm d}(M\dot R)}{{\rm d}t}= 4\pi PR^2,
\end{equation}
where $M$ and $R$ are the mass and the radius of the shell, respectively, $P$
is the pressure, and $\rho_0$ is the ambient density. The above equations can
be rearranged into
\begin{equation}
M = \frac{4\pi}{3}\rho_0R^3\quad\mbox{and}
\end{equation}
\begin{equation}
\ddot R+\frac{3\dot R^2}{R} = \frac{3}{\rho_0}\frac{P}{R}.
\end{equation}
Conservation of energy (total energy of the bubble in the adiabatic phase)
leads to the following equation for the pressure (we adopt an adiabatic index
$\gamma=5/3$)
\begin{equation}
\frac{{\rm d}P}{{\rm d}t}=\frac{\dot E}{2\pi R^3}-5P\frac{\dot
R}{R}+\frac{\rho_0\dot R^3}{R}.
\end{equation}
Instead, in the so-called radiative phase, only the internal energy of the
bubble is conserved, leading to
\begin{equation}
\frac{{\rm d}P}{{\rm d}t}=\frac{\dot E}{2\pi R^3}-5P\frac{\dot R}{R}~.
\end{equation}
In \citet{bp04} the energy was injected just at the initial time by the
supernova, so that $\dot E$ is zero, while here we consider the case of
vanishing initial energy but with a constant $\dot E>0$, which is more
suitable to describe a PWN powered by a long-lived pulsar. By introducing the
quantity
\begin{equation}
y(R)=R^6\dot R^2,\label{eq:y(r)}
\end{equation}
one can transform the equations above into
\begin{equation}
y''(R)=\frac{3\dot E R^5}{\pi\rho_0 \sqrt{y}}+
\left(\frac{6y}{R^2}\right)_{\rm ad},
\label{eq:y''}
\end{equation}
where the term with suffix ``ad'' is added only when treating the adiabatic
phase. One can show that $y(R)=Q^2R^{14/3}$ is a special solution for both the
adiabatic and the radiative equations, with
\begin{equation}
Q_{\rm ad} = \frac{3}{5^{2/3}}\left(\frac{\dot E}{4\pi\rho_0}\right)^{1/3}
\quad\mbox{and} \label{eq:q_ad}
\end{equation}
\begin{equation}
Q_{\rm rad} = \frac{3\cdot2^{1/3}}{77^{1/3}}\left(\frac{\dot
E}{4\pi\rho_0}\right)^{1/3},\label{eq:q_rad}
\end{equation}
respectively. The full transition from the adiabatic to the asymptotic
radiative case requires us to solve (numerically) Equation~(\ref{eq:y''}), by
imposing the continuity of $y(R)$ and $y'(R)$ at the transition. Finally,
given $y(R)$, the time evolution can be computed by solving (numerically also
in this case) Equation~(\ref{eq:y(r)}) for $R(t)$. In the asymptotic cases i.e.\
Equations~(\ref{eq:q_ad}) and (\ref{eq:q_rad}) above, the solutions are:
\begin{equation}
R(t)=\left(\frac{5}{3}Qt\right)^{3/5}.\label{eq:r_asymp}
\end{equation}
\begin{figure}[ht]
\epsscale{1.0}
\centering
\plotone{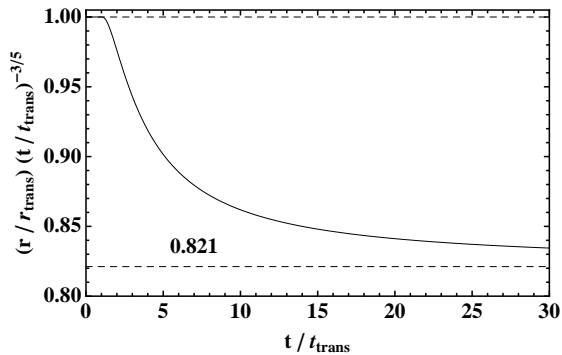}
\caption{Transition from the adiabatic phase to the radiative phase
for a bubble expanding in a uniform medium with continuous energy injection.
The asymptotic behavior in these cases only differs by a small constant factor
of 0.821 \citep[see][and Equations~(\ref{eq:q_ad})--(\ref{eq:r_asymp})]{dok02}.
\label{fig:correction_edot}}
\end{figure}
Therefore, it is apparent that here, apart from a small offset, both the
adiabatic and the asymptotic radiative phases have the same power-law
dependence on time. This is qualitatively different from the case with an
energy release at the initial time, in which the adiabatic (Sedov) phase has
$R(t)\propto t^{2/5}$, while the asymptotic radiative (or ``pressure-driven
snowplow'') phase has $R(t)\propto t^{2/7}$.
This is also different from the behavior of a PWN expanding inside an SNR
\citep[e.g.,][]{rc84}, since we assumed a static ambient medium.
For the adiabatic regime, the
radius estimated here in the thin-layer approximation is only about 10\% smaller
(a factor 0.895) than the true value, as computed by \citet{dok02}; for
the radiative phase, as the layer of matter is thinner, we expect the true
value of the radius to be approximated even better. Using the correct
coefficient for adiabatic evolution, the asymptotic radiative solution is
smaller than the extrapolation of the adiabatic one by a constant factor
0.821. Figure~\ref{fig:correction_edot} shows the radial evolution at
intermediate times. 


\begin{thebibliography}{}
\expandafter\ifx\csname natexlab\endcsname\relax\def\natexlab#1{#1}\fi

\bibitem[{{Acero} {et~al.}(2015){Acero}, {Ackermann}, {Ajello}, {Albert},
  {Atwood}, {Axelsson}, {Baldini}, {Ballet}, {Barbiellini}, {Bastieri},
  {Belfiore}, {Bellazzini}, {Bissaldi}, {Blandford}, {Bloom}, {Bogart},
  {Bonino}, {Bottacini}, {Bregeon}, {Britto}, {Bruel}, {Buehler}, {Burnett},
  {Buson}, {Caliandro}, {Cameron}, {Caputo}, {Caragiulo}, {Caraveo},
  {Casandjian}, {Cavazzuti}, {Charles}, {Chaves}, {Chekhtman}, {Cheung},
  {Chiang}, {Chiaro}, {Ciprini}, {Claus}, {Cohen-Tanugi}, {Cominsky}, {Conrad},
  {Cutini}, {D'Ammando}, {de Angelis}, {DeKlotz}, {de Palma}, {Desiante},
  {Digel}, {Di Venere}, {Drell}, {Dubois}, {Dumora}, {Favuzzi}, {Fegan},
  {Ferrara}, {Finke}, {Franckowiak}, {Fukazawa}, {Funk}, {Fusco}, {Gargano},
  {Gasparrini}, {Giebels}, {Giglietto}, {Giommi}, {Giordano}, {Giroletti},
  {Glanzman}, {Godfrey}, {Grenier}, {Grondin}, {Grove}, {Guillemot}, {Guiriec},
  {Hadasch}, {Harding}, {Hays}, {Hewitt}, {Hill}, {Horan}, {Iafrate}, {Jogler},
  {J{\'o}hannesson}, {Johnson}, {Johnson}, {Johnson}, {Johnson}, {Kamae},
  {Kataoka}, {Katsuta}, {Kuss}, {La Mura}, {Landriu}, {Larsson}, {Latronico},
  {Lemoine-Goumard}, {Li}, {Li}, {Longo}, {Loparco}, {Lott}, {Lovellette},
  {Lubrano}, {Madejski}, {Massaro}, {Mayer}, {Mazziotta}, {McEnery},
  {Michelson}, {Mirabal}, {Mizuno}, {Moiseev}, {Mongelli}, {Monzani},
  {Morselli}, {Moskalenko}, {Murgia}, {Nuss}, {Ohno}, {Ohsugi}, {Omodei},
  {Orienti}, {Orlando}, {Ormes}, {Paneque}, {Panetta}, {Perkins},
  {Pesce-Rollins}, {Piron}, {Pivato}, {Porter}, {Racusin}, {Rando}, {Razzano},
  {Razzaque}, {Reimer}, {Reimer}, {Reposeur}, {Rochester}, {Romani},
  {Salvetti}, {S{\'a}nchez-Conde}, {Saz Parkinson}, {Schulz}, {Siskind},
  {Smith}, {Spada}, {Spandre}, {Spinelli}, {Stephens}, {Strong}, {Suson},
  {Takahashi}, {Takahashi}, {Tanaka}, {Thayer}, {Thayer}, {Thompson},
  {Tibaldo}, {Tibolla}, {Torres}, {Torresi}, {Tosti}, {Troja}, {Van Klaveren},
  {Vianello}, {Winer}, {Wood}, {Wood}, {Zimmer}, \& {Fermi-LAT
  Collaboration}}]{aaa+15}
{Acero}, F., {Ackermann}, M., {Ajello}, M., {et~al.} 2015, \apjs, 218, 23

\bibitem[{{Bandiera} \& {Petruk}(2004)}]{bp04}
{Bandiera}, R., \& {Petruk}, O. 2004, \aap, 419, 419

\bibitem[{{Bandiera} \& {Petruk}(2010)}]{bp10}
---. 2010, \aap, 509, A34

\bibitem[{{Bandiera} \& {Petruk}(2016)}]{bp16}
---. 2016, \mnras, 459, 178

\bibitem[{{Ben Bekhti} {et~al.}(2016){Ben Bekhti}, {Fl{\"o}er}, {Keller},
  {Kerp}, {Lenz}, {Winkel}, {Bailin}, {Calabretta}, {Dedes}, {Ford}, {Gibson},
  {Haud}, {Janowiecki}, {Kalberla}, {Lockman}, {McClure-Griffiths}, {Murphy},
  {Nakanishi}, {Pisano}, \& {Staveley-Smith}}]{bfk+16}
{Ben Bekhti}, N., {Fl{\"o}er}, L., {Keller}, R., {et~al.} 2016, \aap, 594, A116

\bibitem[{{Bock} {et~al.}(1999){Bock}, {Large}, \& {Sadler}}]{bls99}
{Bock}, D.~C.-J., {Large}, M.~I., \& {Sadler}, E.~M. 1999, \aj, 117, 1578

\bibitem[{{Briggs}(1995)}]{bri95}
{Briggs}, D.~S. 1995, \baas, 27, 1444

\bibitem[{{Campbell-Wilson} \& {Hunstead}(1994)}]{ch94}
{Campbell-Wilson}, D., \& {Hunstead}, R.~W. 1994, \pasa, 11, 33

\bibitem[{{Cordes} \& {Lazio}(2002)}]{cl02}
{Cordes}, J.~M., \& {Lazio}, T.~J.~W. 2002, astro-ph/0207156

\bibitem[{{Cordes} {et~al.}(1993){Cordes}, {Romani}, \& {Lundgren}}]{crl93}
{Cordes}, J.~M., {Romani}, R.~W., \& {Lundgren}, S.~C. 1993, \nat, 362, 133

\bibitem[{{Dame} {et~al.}(2001){Dame}, {Hartmann}, \& {Thaddeus}}]{dht01}
{Dame}, T.~M., {Hartmann}, D., \& {Thaddeus}, P. 2001, \apj, 547, 792

\bibitem[{{Dokuchaev}(2002)}]{dok02}
{Dokuchaev}, V.~I. 2002, \aap, 395, 1023

\bibitem[{{Dubner} \& {Giacani}(2015)}]{dg15}
{Dubner}, G., \& {Giacani}, E. 2015, \aapr, 23, 3

\bibitem[{{Dutra} \& {Bica}(2002)}]{db02}
{Dutra}, C.~M., \& {Bica}, E. 2002, \aap, 383, 631

\bibitem[{{Feitzinger} \& {Stuewe}(1984)}]{fs84}
{Feitzinger}, J.~V., \& {Stuewe}, J.~A. 1984, \aaps, 58, 365

\bibitem[{{Ferri{\`e}re}(2001)}]{fer01}
{Ferri{\`e}re}, K.~M. 2001, Reviews of Modern Physics, 73, 1031

\bibitem[{{Fitzpatrick}(2004)}]{fit04}
{Fitzpatrick}, E.~L. 2004, in Ast.\ Soc.\ of the Pac.\ Conference
Series,  309, Astrophysics of Dust, ed. A.~N.
  {Witt}, G.~C. {Clayton}, \& B.~T. {Draine}, 33

\bibitem[{{Frail} {et~al.}(1994){Frail}, {Kassim}, \& {Weiler}}]{fkw94}
{Frail}, D.~A., {Kassim}, N.~E., \& {Weiler}, K.~W. 1994, \aj, 107, 1120

\bibitem[{{Gaensler} \& {Slane}(2006)}]{gs06}
{Gaensler}, B.~M., \& {Slane}, P.~O. 2006, \araa, 44, 17

\bibitem[{{Graham}(1970)}]{gra70}
{Graham}, J.~A. 1970, \aj, 75, 703

\bibitem[{{Green} {et~al.}(1999){Green}, {Cram}, {Large}, \& {Ye}}]{gcl+99}
{Green}, A.~J., {Cram}, L.~E., {Large}, M.~I., \& {Ye}, T. 1999, \apjs, 122,
  207

\bibitem[{{Green} {et~al.}(2014){Green}, {Reeves}, \& {Murphy}}]{grm14}
{Green}, A.~J., {Reeves}, S.~N., \& {Murphy}, T. 2014, \pasa, 31, e042

\bibitem[{{G{\"u}ver} \& {{\"O}zel}(2009)}]{go09}
{G{\"u}ver}, T., \& {{\"O}zel}, F. 2009, \mnras, 400, 2050

\bibitem[{{Haverkorn} {et~al.}(2006){Haverkorn}, {Gaensler},
  {McClure-Griffiths}, {Dickey}, \& {Green}}]{hgm+06}
{Haverkorn}, M., {Gaensler}, B.~M., {McClure-Griffiths}, N.~M., {Dickey},
  J.~M., \& {Green}, A.~J. 2006, \apjs, 167, 230

\bibitem[{{H{\"o}gbom}(1974)}]{hog74}
{H{\"o}gbom}, J.~A. 1974, \aaps, 15, 417

\bibitem[{{Johnston} {et~al.}(2005){Johnston}, {Hobbs}, {Vigeland}, {Kramer},
  {Weisberg}, \& {Lyne}}]{jhv+05}
{Johnston}, S., {Hobbs}, G., {Vigeland}, S., {et~al.} 2005, \mnras, 364, 1397

\bibitem[{{Johnston} \& {Weisberg}(2006)}]{jw06}
{Johnston}, S., \& {Weisberg}, J.~M. 2006, \mnras, 368, 1856

\bibitem[{{Kargaltsev} \& {Pavlov}(2008)}]{kp08}
{Kargaltsev}, O., \& {Pavlov}, G.~G. 2008, in AIP Conf. Proc.,  983, 40 Years
  of Pulsars: Millisecond Pulsars, Magnetars and More, ed. C.~{Bassa},
  Z.~{Wang}, A.~{Cumming}, \& V.~M. {Kaspi} (Melville, NY: AIP), 171

\bibitem[{{Klingler} {et~al.}(2016){Klingler}, {Kargaltsev}, {Rangelov},
  {Pavlov}, {Posselt}, \& {Ng}}]{kkr+16}
{Klingler}, N., {Kargaltsev}, O., {Rangelov}, B., {et~al.} 2016, \apj, 828, 70

\bibitem[{{Kothes} \& {Brown}(2009)}]{kb09}
{Kothes}, R., \& {Brown}, J.-A. 2009, in IAU Symp.,  259, Cosmic Magnetic
  Fields: From Planets, to Stars and Galaxies, ed. K.~G. {Strassmeier}, A.~G.
  {Kosovichev}, \& J.~E. {Beckman} (New York: Cambridge Univ.\ Press), 75

\bibitem[{{Kramer} {et~al.}(2003){Kramer}, {Bell}, {Manchester}, {Lyne},
  {Camilo}, {Stairs}, {D'Amico}, {Kaspi}, {Hobbs}, {Morris}, {Crawford},
  {Possenti}, {Joshi}, {McLaughlin}, {Lorimer}, \& {Faulkner}}]{kbm+03}
{Kramer}, M., {Bell}, J.~F., {Manchester}, R.~N., {et~al.} 2003, \mnras, 342,
  1299

\bibitem[{{Ma} {et~al.}(2016){Ma}, {Ng}, {Bucciantini}, {Slane}, {Gaensler}, \&
  {Temim}}]{mnb+16}
{Ma}, Y.~K., {Ng}, C.-Y., {Bucciantini}, N., {et~al.} 2016, \apj, 820, 100

\bibitem[{{Misanovic} {et~al.}(2002){Misanovic}, {Cram}, \& {Green}}]{mcg02}
{Misanovic}, Z., {Cram}, L., \& {Green}, A. 2002, \mnras, 335, 114

\bibitem[{{Morlino} {et~al.}(2015){Morlino}, {Lyutikov}, \& {Vorster}}]{mlv15}
{Morlino}, G., {Lyutikov}, M., \& {Vorster}, M. 2015, \mnras, 454, 3886

\bibitem[{{Murphy} {et~al.}(2007){Murphy}, {Mauch}, {Green}, {Hunstead},
  {Piestrzynska}, {Kels}, \& {Sztajer}}]{mmg+07}
{Murphy}, T., {Mauch}, T., {Green}, A., {et~al.} 2007, \mnras, 382, 382

\bibitem[{{Ng} {et~al.}(2012){Ng}, {Bucciantini}, {Gaensler}, {Camilo},
  {Chatterjee}, \& {Bouchard}}]{nbg+12}
{Ng}, C.-Y., {Bucciantini}, N., {Gaensler}, B.~M., {et~al.} 2012, \apj, 746,
  105

\bibitem[{{Ng} {et~al.}(2010){Ng}, {Gaensler}, {Chatterjee}, \&
  {Johnston}}]{ngc+10}
{Ng}, C.-Y., {Gaensler}, B.~M., {Chatterjee}, S., \& {Johnston}, S. 2010, \apj,
  712, 596

\bibitem[{{Ng} \& {Romani}(2007)}]{nr07}
{Ng}, C.-Y., \& {Romani}, R.~W. 2007, \apj, 660, 1357

\bibitem[{{Noutsos} {et~al.}(2012){Noutsos}, {Kramer}, {Carr}, \&
  {Johnston}}]{nkc+12}
{Noutsos}, A., {Kramer}, M., {Carr}, P., \& {Johnston}, S. 2012, \mnras, 423,
  2736

\bibitem[{{Pacholczyk}(1970)}]{pac70}
{Pacholczyk}, A.~G. 1970, {Radio Astrophysics. Nonthermal Processes in Galactic
  and Extragalactic Sources} (San Francisco: Freeman)

\bibitem[{{Reynolds}(1998)}]{rey98}
{Reynolds}, S.~P. 1998, \apj, 493, 375

\bibitem[{{Reynolds} \& {Chevalier}(1984)}]{rc84}
{Reynolds}, S.~P., \& {Chevalier}, R.~A. 1984, \apj, 278, 630

\bibitem[{{Romanova} {et~al.}(2005){Romanova}, {Chulsky}, \&
  {Lovelace}}]{rcl05}
{Romanova}, M.~M., {Chulsky}, G.~A., \& {Lovelace}, R.~V.~E. 2005, \apj, 630,
  1020

\bibitem[{{Sault} {et~al.}(1999){Sault}, {Bock}, \& {Duncan}}]{sbd99}
{Sault}, R.~J., {Bock}, D.~C.-J., \& {Duncan}, A.~R. 1999, \aaps, 139, 387

\bibitem[{{Sault} {et~al.}(1995){Sault}, {Teuben}, \& {Wright}}]{stw95}
{Sault}, R.~J., {Teuben}, P.~J., \& {Wright}, M.~C.~H. 1995, in ASP Conf. Ser.,
  ~77, Astronomical Data Analysis Software and Systems IV, ed. R.~A. {Shaw},
  H.~E. {Payne}, \& J.~J.~E. {Hayes} (San Francisco, CA: ASP), 433

\bibitem[{{Temim} {et~al.}(2015){Temim}, {Slane}, {Kolb}, {Blondin}, {Hughes},
  \& {Bucciantini}}]{tsk+15}
{Temim}, T., {Slane}, P., {Kolb}, C., {et~al.} 2015, \apj, 808, 100

\bibitem[{{Torres} {et~al.}(2001){Torres}, {Butt}, \& {Camilo}}]{tbc01}
{Torres}, D.~F., {Butt}, Y.~M., \& {Camilo}, F. 2001, \apjl, 560, L155

\bibitem[{{Truelove} \& {McKee}(1999)}]{tm99}
{Truelove}, J.~K., \& {McKee}, C.~F. 1999, \apjs, 120, 299

\bibitem[{{van Kerkwijk} \& {Ingle}(2008)}]{vi08}
{van Kerkwijk}, M.~H., \& {Ingle}, A. 2008, \apjl, 683, L159

\bibitem[{{Wang} {et~al.}(2014){Wang}, {Ng}, {Wang}, {Li}, \&
  {Kaplan}}]{wnw+14}
{Wang}, Z., {Ng}, C.-Y., {Wang}, X., {Li}, A., \& {Kaplan}, D.~L. 2014, \apj,
  793, 89

\bibitem[{{Wilkin}(1996)}]{wil96}
{Wilkin}, F.~P. 1996, \apjl, 459, L31

\bibitem[{{Williams} {et~al.}(2015){Williams}, {Rangelov}, {Kargaltsev}, \&
  {Pavlov}}]{wrk+15}
{Williams}, B.~J., {Rangelov}, B., {Kargaltsev}, O., \& {Pavlov}, G.~G. 2015,
  \apjl, 808, L19

\bibitem[{{Wilson} {et~al.}(2011){Wilson}, {Ferris}, {Axtens}, {Brown},
  {Davis}, {Hampson}, {Leach}, {Roberts}, {Saunders}, {Koribalski}, {Caswell},
  {Lenc}, {Stevens}, {Voronkov}, {Wieringa}, {Brooks}, {Edwards}, {Ekers},
  {Emonts}, {Hindson}, {Johnston}, {Maddison}, {Mahony}, {Malu}, {Massardi},
  {Mao}, {McConnell}, {Norris}, {Schnitzeler}, {Subrahmanyan}, {Urquhart},
  {Thompson}, \& {Wark}}]{wfa+11}
{Wilson}, W.~E., {Ferris}, R.~H., {Axtens}, P., {et~al.} 2011, \mnras, 416, 832

\bibitem[{{Yusef-Zadeh} \& {Bally}(1987)}]{yb87}
{Yusef-Zadeh}, F., \& {Bally}, J. 1987, \nat, 330, 455

\bibitem[{{Yusef-Zadeh} \& {Gaensler}(2005)}]{yg05}
{Yusef-Zadeh}, F., \& {Gaensler}, B.~M. 2005, Advances in Space Research, 35,
  1129

\end{thebibliography}

\end{document}